\documentclass[journal,twoside]{IEEEtran}
\usepackage{pgfplots} 
\usepackage{tikz}
\usepackage{graphicx,cite,multirow,rotating}
\usepackage{amssymb}
\usepackage{amsfonts}
\usepackage{yfonts}
\usepackage{psfrag}
\usepackage{amsfonts}
\usepackage{algorithm}
\usepackage{algorithmic}
\usepackage{lipsum}
\usepackage{bbm}
\usepackage{graphicx}
\usepackage[caption=false]{subfig}
\usepackage{dsfont}

\def\rE{{\rm E}}

\def\arg{{\rm arg}}

\def\d={{\stackrel {\rm def}{=}}}

\newcommand{\beq}{\begin{equation}}

\newcommand{\eeq}{\end{equation}}
\newcommand{\bqy}{\begin{eqnarray}}
\newcommand{\eqy}{\end{eqnarray}}

\newcommand{\xx}[1]{{\mathbf x}}
\newcommand{\yy}[1]{{\mathbf{y}}}
\newcommand{\XX}[1]{{\mathbf{X}}}
\newcommand{\YY}[1]{{\mathbf{Y}}}
\newcommand{\RR}[1]{{\mathbf{R}}}
\newcommand{\HH}[1]{{\mathcal{H}}}

\def\BD{\begin{displaymath}}
\def\BEA{\begin{eqnarray}}
\def\BEAs{\begin{eqnarray*}}
\def\ED{\end{displaymath}}
\def\EE{\end{equation}}
\def\EEA{\end{eqnarray}}
\def\EEAs{\end{eqnarray*}}

\def\bH{{\bf H}}
\def\bh{{\bf h}}
\def\bI{{\bf I}}

\def\bn{{\bf n}}

\def\bR{{\bf R}}
\def\br{{\bf r}}
\def\bS{{\bf S}}
\def\bs{{\bf s}}
\def\bT{{\bf T}}

\def\bW{{\bf W}}

\def\bX{{\bf X}}
\def\bx{{\bf x}}

\def\by{{\bf y}}

\def\rA{{\rm A}}

\def\rB{{\rm B}}

\def\rC{{\rm C}}

\def\rE{{\rm E}}

\def\rH{{\rm H}}

\def\rP{{\rm P}}

\def\rq{{\rm q}}

\def\rr{{\rm r}}

\def\rT{{\rm T}}

\def\rv{{\rm v}}

\def\b_eta{\mbox{\boldmath $\eta$}}

\def\bnu{\mbox{\boldmat $\nu$}}

\def\L2{L^{2}[-{\pi \over 2} , {\pi \over 2}]}

\def\bnu{\bnu}

\usepackage{setspace}
\usepackage{color}
\usepackage{amsmath,stackrel,dsfont}
\usepackage{amssymb,hyperref,stmaryrd}
\pgfplotsset{width=10cm,compat=newest}
\usepgfplotslibrary{units}
\usetikzlibrary{spy,backgrounds}

{}

\def\b1{{\mathbf{1}}}
\def\rE{{\rm E}}

\title{  {Lossless Size Reduction for Integer Least Squares with Application to Sphere Decoding}}

\author{Mohammad~Neinavaie, ~Mostafa~Derakhtian and Sergiy A. Vorobyov, IEEE Fellow
	\thanks{M. Neinavaie and S. A. Vorobyov are with Dept.  of  Signal  Processing  and  Acoustics, Aalto  University, Aalto, Finland.
		 M. Derakhtian  is with the Dept. of Electrical and Computer Engineering, Shiraz University, Shiraz, Iran. (email: \{neinavaie. mohammad,Sergiy.vorobyov\}@aalto.fi, derakhtian@shirazu.ac.ir).
}
}
\begin{document}

\maketitle

\begin{abstract}
 Minimum achievable complexity (MAC) for a maximum likelihood (ML) performance-achieving detection algorithm is derived. Using the derived MAC, we prove that the conventional sphere decoding (SD) algorithms suffer from an inherent weakness at low SNRs. To find a solution for the low SNR deficiency,  we analyze  the effect of  zero-forcing (ZF) and minimum mean square error (MMSE) detected symbols on the MAC and demonstrate that although they both improve the SD algorithm in terms of the computational complexity, the MMSE point has  a vital difference at low SNRs. By  exploiting the information provided by the MMSE method, we prove the existence of a lossless size reduction which can be interpreted as the feasibility of a detection  method which is capable of detecting the ML symbol  without visiting any nodes at low and high SNRs. We also propose a lossless size reduction-aided detection method which  achieves the promised complexity bounds marginally and reduces the overall computational complexity significantly, while obtaining the ML performance. The theoretical analysis is corroborated with numerical simulations.  
\end{abstract}
\section{Introduction} \label{introduction}
\IEEEPARstart{W}HEN  the transmitted symbols are from a finite set, the problem of optimal detection is an integer least-squares (ILS) problem. It arises in many applications such as code division multiple access (CDMA) systems \cite{hasibi5},  Vertical Bell Labs
Layered Space-Time (V-BLAST) structure \cite{Tse}, linear dispersion space-time  block coding (LD-STBC)  \cite{jalden4} and generalized spatial modulation (GSM) schemes \cite{gsm} to name just a few most frequent and established applications.  The maximum likelihood (ML) detector results in an optimum solution of the ILS problem. However, the ML detection problem can be solved by a brute-force search which is  computationally infeasible \cite{tsebook}. Thus, sphere-decoding (SD) was proposed as an efficient tree-search based method to obtain the ML solution \cite{50years}. Unlike the brute force search, the SD algorithm searches over the lattice points that lie within a hyper sphere of radius $R$ around the received signal. The SD algorithm was first introduced by Fincke and Pohst (FP) in \cite{finke}. Later, a more efficient variation of the SD algorithm known as Schnor-Euchner (SE) refinement was presented in \cite{Schnorr}. Based on FP and SE algorithms, two improvements of these variations of the SD algorithm were proposed in \cite{Damen}, and shown to offer a computational complexity  reduction compared to the SE and FP variations.

The  computational complexity of the SD algorithm was studied in  \cite{hasibi}-\hspace*{-5px}\cite{jalden}. In \cite{hasibi} and \cite{hasibicomp}, the expected number of operations required by the algorithm was derived over the Rayleigh fading channel, realization of the noise and transmitted symbols. It was shown that the expected complexity has a polynomial behavior at a wide range of signal-to-noise ratios (SNRs). Contrasted with this claim, the main result of  \cite{jalden} is that the expected complexity of the SD is exponential for a fixed SNR, and the complexity exponent behavior  shows that the SD algorithm is not efficient for the systems which operate under low-SNR conditions or have a large size. In \cite{jaldeninf}, the complexity distribution for random infinite lattices for the SD algorithm was also analyzed. With a focus on the space-time codes (STC) and diversity-multiplexing tradeoff (DMT)  optimality, the minimum complexity for the SD algorithm, which achieves a vanishing gap to the ML performance at high SNRs, was derived in \cite{jaldenexp}. This minimum complexity is described via introducing the SD complexity exponent. Aiming for reducing the computational complexity, there are a number of other variants of the SD algorithm which can be divided according to the error performance  into optimal and suboptimal SD algorithms  \cite{13vblast}-\hspace*{-3.5px}\cite{widely_linear}.         

At the opposite extreme,  linear detection (LD) schemes including the zero-forcing  (ZF)  and minimum mean square error (MMSE), perform linear operations which incur a much lower complexity compared to the SD algorithm \cite{tsebook}. This polynomial complexity comes at the cost of a dramatic performance loss\cite{zeroforcing}. As an example of auxiliary/initial, low cost information, the LD  method sometimes results in a reduced search space for the SD algorithm \cite{Ma}-\hspace*{-3.5px}\cite{tradeoff}. Most of these schemes  achieve either a sub-optimal performance to reduce the computational complexity, e.g., \cite{2012}, or have exponential computational complexity to achieve the ML performance, e.g.,  \cite{2017_16}.  In \cite{neinavaie}, an ML performance-achieving   method  which is as complex as the ZF method with probability one, at high SNRs was proposed. In other words,  at a high SNR regime, this method meets both complexity and performance extremes with probability one. However, this method does not achieve the optimal performance at a low SNR regime and only operates in a V-BLAST system with the Rayleigh fading channel.

A fundamental question for SD algorithms is the {\it minimum possible search space}  to achieve the optimal ML performance when auxiliary low cost information, such as the  detected symbol by the LD method or an initial radius, is available. In this paper, we address this question by presenting the minimum achievable complexity (MAC) of an SD algorithm. Moreover, the exponential behavior of the SD algorithm at low SNRs and large number of input symbols is, arguably, the most significant drawbacks of SD algorithm \cite{jalden}. In this paper, theoretical study and possible solutions to overcome these drawbacks are given.  
\vspace*{-3.5mm}
\subsection{Main Contributions}
  1) Using the law of large numbers, we obtain the minimum feasible complexity, i.e., MAC of the SD algorithm for a given set of initial information. This set may solely contain the classic initial information of the conventional SD algorithms, e.g., the initial radius and the channel matrix, or it may involve more information, such as the initial point detected by an LD method, hereafter denoted by $\bs_{\rm LD}$. This  complexity bound can be used  as  a bench mark to determine how well a particular SD algorithm exploits the initial information.  The proposed complexity bound, MAC, is distinct from the complexity exponent  in \cite{jaldenexp} in some important respects. Unlike the complexity exponent, which only considers the high SNR optimal performance achieving SD, the MAC is the minimum  complexity bound for the SD algorithms which achieve the exact ML performance at all SNRs. Moreover, our bound describes the complexity behavior in a more general sense. Indeed, it is not limited to a specific setting or a type of fading channel. 

 2) The proposed MAC reveals an inherent limitation of some conventional SD algorithms at a low SNR regime. More precisely, we prove that the results obtained in \cite{jalden} about the weakness of the SD algorithm at small SNRs are natural and the auxiliary information of these SD algorithms is not effective at low SNRs.    

 3) Although, the size reduction notion, in the sense that SD algorithms can search over a reduced search space with a performance loss,  already exists in the literature, e.g., \cite{2012}, it is in this paper that the concept of a {\it lossless size reduction} in SD algorithms is introduced as the capability of reducing the effective size without any performance loss.  Along with introducing  the concept of a lossless size reduction, the derived MAC allows us to prove that by adding {\it appropriate low cost auxiliary information},   a solution for the low SNR deficiency problem of conventional SD algorithms is attainable. Specifically, we show that, unlike the ZF detected symbol, adding  the MMSE detected symbol to the information set helps to overcome the weakness of conventional SD algorithms at low SNRs.  

 4) In addition to capabilities and bounds, an ML performance-achieving SD algorithm which follows the lossless size reduction concept is proposed. We prove that this algorithm intelligently deploys the initial information to reduce the effective problem size without any performance loss. It is also capable of operating under any type of fading channels.   By defining marginal optimality as achieving the optimal error performance without any tree-search at low and high SNRs, we prove that the proposed algorithm is marginally optimal.   The distinct characteristics of the proposed algorithm which, to the best of our knowledge, are not met in any other SD algorithms are as follows: providing a lossless size reduction, i.e., reducing the effective size while being optimal in the sense of error performance; an exemption from the tree-search at a range of SNRs, especially low SNRs.

The remainder of the paper is organized as follows. After  reviewing the basic SD algorithm and its computational complexity  in  Section \ref{sphere}, we derive the MAC of SD algorithms in Section \ref{mac}. We present the lossless size reduction concept  in Section \ref{llsrsection} and propose the marginally optimal SD algorithm  in Section \ref{mcosd}. Section \ref{simulation} explores the validity of theoretical results via simulations. Finally, Section \ref{conclusion} concludes  the paper.
\vspace*{-3.5mm}
\subsection{Notations}
Throughout the paper boldface small letters denote vectors and boldface capital letters denote matrices. A superscript $(\cdot)^T$ denotes the transpose of a matrix,  $|x|$ stands for  the absolute value of $x$ and $\|\bx\|$ stands for the  Euclidean norm, $[\bX]_{mn}$ denotes the element in the $m$th row and the $n$th column of matrix $\bX$ and $[\bx]_i$ is the $i$th element of  vector $\bx$. The probability of event $\rA$  is denoted as $\rP(\rA)$. Mathematical expectation of $X$ over variable $Y$ is denoted by $\rE_{Y}\{X\}$, while $f(X)$ is the probability density function (pdf) of the random variable $X$ and $f(X|Y)$ is the conditional pdf of the random variable $X$ given $Y$. We also use the notation $\dot{=}$ to denote the asymptotic exponential equality, i.e., $f(\rho)\dot{=}g(\rho)$ means that $\lim \limits_{\rho \rightarrow \infty}\frac{\log f(\rho)}{\log\rho}=\lim \limits_{\rho \rightarrow \infty}\frac{\log g(\rho)}{\log\rho}$.  
 Finally, $\mathds{C}$ denotes the field of complex numbers and $|\mathcal{S}|$ is the cardinality of set $\mathcal{S}$. 
\section{Sphere Decoding}\label{sphere}
 Consider the  system model
\beq \label{systemmodel}
\by = \bH \bs + \bn,
\eeq
where $\bH$ is the $L\times K$  channel matrix, the $K\times 1$ vector $\bs \in  \mathcal{C}^K$ consists of the information symbols drawn from an arbitrary $M$-ary quadrature amplitude modulation (QAM) or pulse amplitude modulation (PAM) constellation,  $\by$ denotes the  $L\times 1$ received vector, and $\bn$ is the independent and identically distributed (i.i.d.) additive complex Gaussian noise with   variance  $\sigma^2$, and the SNR is defined as ${\rm \rho}\triangleq \frac{1}{\sigma^2}$. The channel $\bH$ is general enough to capture a number of applications such as, uncoded V-BLAST and STC MIMO systems,  precoded orthogonal frequency-division multiplexing (OFDM) systems \cite{MA19}, and multiuser systems \cite{MA35}.

The ML solution for the information symbol decoding is 
\beq \label{problem}
{\bs}_{\rm ML} = \arg\min_{\bs\in \mathcal{C}^K}\|\by -\bH\bs\|^2,
\eeq
which is the closest lattice point search problem. Unlike the brute-force ML detector, the SD  algorithm reduces the computational complexity by searching over the lattice points inside a sphere with radius $R$.  The essence of any SD-based method is the FP algorithm and its refinements, such as the SE enumeration \cite{50years}.  
 
The computational complexity of the SD algorithm is a function of the number of visited nodes in the tree-search \cite{hasibi}. In this paper, as in \cite{jalden} and \cite{jaldeninf}, we consider the number of visited nodes itself as the complexity measure. The number of visited nodes in the SD algorithm is equal to the sum of visited lattice points at the $k$th layer, $k=1,\dots,K$. If we denote the number of lattices, in a $k$ dimensional hypersphere with radius $R$, by ${N}_k$, the number of visited nodes of the SD algorithm is \cite{jaldeninf} 
\beq \label{AA}
{\bar{N}}_{\rm SD} = \sum_{k=1}^K \rE\{N_k\},
\eeq
where 
$
{N}_k= \left|\left\{  \bs^k \in \mathcal{C}^k \left| \|\tilde{\by}^k-\bR_k \bs^k\|^2\leq R   \right.\right\}\right|.
$
Indeed, $N_k$ depends on the information set, denoted by $\mathcal{I}$, which is fed to the algorithm. This initial information is mostly embedded in the preprocessing stage and the initial choice of $R$.   The preprocessing stage is  comprised of  the QR decomposition of the channel matrix $\bH$, and for some variants of the SD algorithm, also  an LD detected symbol, e.g., the ZF or MMSE points, and sometimes an ordering stage. The ordering  can be a natural back-substitution or a more sophisticated one which is topically based on the channel matrix realization $\bH$. For example, see the ordering algorithms in \cite{Damen}.  For the initial radius $R$, the two most well-known choices are {\it lattice-independent} and {\it lattice-dependent} radii \cite{jaldeninf}. The FP variant of the SD method primarily adopts a lattice-independent radius, and the SE refinement is a lattice-dependent based SD algorithm.    
A powerful choice for a lattice-independent  radius is proposed in \cite[Section 4]{hasibi}.

 The information set for the  SD algorithm presented in \cite{hasibi} is the channel matrix and an initial radius, i.e., $\mathcal{I}_{\rm li} = \left\{ \bH , R_{\rm H} \right\}$. The initial radius  presented in \cite{hasibi}, denoted by $R_{\rm H}$, guarantees the existence of at least one lattice point inside the sphere with a high probability.  In the lattice-dependent variants of  the SD algorithm, such as the SE refinement, the provided information set is $\mathcal{I}_{\rm ld} =\{\bH,R_\rB \}$, where  $R_{\rB}$  is the initial  lattice-dependent radius obtained by $R_\rB = \| \by-\bH\bs_{\rB}\|^2$, and $\bs_{\rB}$ is the Babai point or the zero forcing-decision feedback (ZF-DF) estimate \cite{Schnorr}. 
 
 It is worth mentioning that the ZF-DF point can be modified to the other points, such as the MMSE detected symbol. Intuitively, the more information the additional point gives us about the ML detected symbol, the more effective the initial information set becomes. Therefore, if we add a point which is very likely to be the ML detected symbol, the set $\mathcal{I}$ plays a more important role in reducing the computational complexity.  Nevertheless, this additional auxiliary point should not impose a high computational complexity to the SD algorithm itself. This is the reason that traditional SD algorithms commonly deploy LD methods which incur an acceptably low, often polynomial, complexity.  
\section{MAC in an SD algorithm} \label{mac}
As it was mentioned previously, the initial information, $\mathcal{I}$, can potentially reduce the computational complexity of the SD algorithm. In other words, the initial information may make it unnecessary to search over all the lattice points inside the sphere. As an extreme example, assume that $\mathcal{I} = \left\{ \bs_\rB \right\}$ and we know that $\bs_\rB$ is the ML detected symbol with probability one. Therefore, we lose nothing, with probability one, if we only visit the nodes corresponding to $\bs_\rB$. Thus, in general,  according to this information, some of the lattice points become  probable or {\it typical}. The typical set, or equivalently, the set of  minimum possible lattice points,  can be determined  using the asymptotic equipartition property (AEP)  and the law of large numbers \cite{gallager}.  If the typical set of lattice points is provided, the SD algorithm only needs to perform the search over this set. The typical set concept allows us to obtain the minimum necessary and possible  number of lattice points. Then, an algorithm which searches over the typical set is able to find the ML solution with the minimum possible number of visited nodes. Hereafter, this minimum possible number of visited nodes is referred to as the MAC of an SD algorithm which can be interpreted as a complexity bound for an SD algorithm. Achieving  this  bound for the number of visited nodes is realizable by constructing the typical set and searching over it,  and can be considered as a benchmark for how efficient a variant of an SD algorithm is, in terms of computational complexity. By a marginal analysis of this bound, we investigate the inherent drawback of the conventional variants of the SD algorithm which appear primarily at the low SNR regime. The basic difference between this type of analysis  and the lower bounds obtained in the literature, for instance  \cite{jalden}, should be noticed carefully. 

In the SD complexity analysis presented in \cite{jalden}, the lower bounds are derived for a specific SD method and show the limitations of those methods, exclusively, with a certain radius, body search strategy, and preprocessing method. Nonetheless, the MAC, obtained by the law of large numbers, reveals the smallest possible search space for any given variant of the SD algorithm and, therefore,  the inherent limitations of the SD algorithm with a given set of information is clarified.
Indeed, the MAC is the {\it ultimate} potential of an SD algorithm. 

To obtain this lower bound, we consider a sequence of $N$ ML search candidates as
\beq
\bS_{\rm ML}^k = [\bs^k_{\rm ML}[1],\bs^k_{\rm ML}[2],\dots,\bs^k_{\rm ML}[N]]^T,
\eeq
where $\bs^k_{\rm ML}[n]$ is the $k$ dimensional ML search candidate at the $n$th time slot. To obtain the possible sequences $\bS_{\rm ML}^k$ or the typical sets, the probability $\rP(\bs_{\rm ML}^k|\mathcal{I})$ needs to be calculated. By looking at large sequences of these symbols and using the law of large numbers, these probabilities reflect on  which  possible sequences of  the  ML search candidates $\bs_{\rm ML}^{k}$'s are more likely to happen. As in Shannon's  AEP, the set of these probable sequences for a large number of observations $N$ is referred to as the  typical set \cite{gallager}. 
 According to the law of large numbers, if the typical set has to be available, we can find the ML solution by searching over the typical set.
We also know that the sequences of the typical set are equiprobable, which means that if we search over a smaller set than the typical set, we will miss ${\bs}_{\rm ML}$ with a considerable probability.  Therefore,   in order to find the ML point, the typical set has to be the minimum sufficient and necessary search space. In the following lemma, we present the MAC, denoted by $\rC^{min}_{\rm SD}$,   for a given information set $\mathcal{I}$.

 {\it Lemma 1:} The MAC of a given SD algorithm with the initial information set $\mathcal{I}$ is  
 \beq \label{car}
 \rC^{min}_{\rm SD} = \rE_{\mathcal{I}} \left\{\sum_{k=1}^KM^{-\rE_{\bs^k_{\rm ML}}\left\{ \log\rP\left(\bs_{\rm ML}^k|\mathcal{I}\right)\right\}}\right\}.
 \eeq
 
 {\it Proof:} See Appendix \ref{lemma1proof}.
 
 In (\ref{car}), $ -\rE_{\bs^k_{\rm ML}}\left\{ \log\rP\left(\bs_{\rm ML}^k|\mathcal{I}\right)\right\}$ is the entropy of  $\bs_{\rm ML}^k$  given the set $\mathcal{I}$. Designing a method which solely searches over the typical set, and hence achieves the MAC, is attainable, yet seems to be a challenging problem. Nonetheless, the MAC can be considered as a computational complexity bench mark. It should be noted that the MAC is only a function of the initial information set, i.e.,  $\mathcal{I}$. As it was mentioned previously, according to the AEP and  for a given initial information set,  no matter which radius, body search strategy or preprocessing is adopted, if the number of visited  nodes is smaller than $\rC^{min}_{\rm SD}$, the ML performance is not achieved.   

  The following theorem reveals a fundamental low SNR limitation of the conventional SD methods with  lattice-independent and lattice-dependent radii. It should be noted that in the proofs  concerning  the lattice-independent radii, we have considered an FP method with the lattice-independent  radius in \cite{hasibi} which is also adopted as the radius of the SD algorithm in many other FP algorithms, e.g., \cite{jaldeninf}. Also, for the lattice-dependent based SD algorithm,  we have considered an SD method which applies the lattice-dependent radius $R_\rB = \|\by-\bH\bs_{\rm LD}\|^2$. Moreover, regimes with $\rho\rightarrow 0$ and $\rho \rightarrow \infty$ are sometimes referred to as the low and high SNR regimes, respectively. 
  
 {\it Theorem 1:} The MAC of a lattice-independent and lattice-dependent conventional SD algorithm satisfies the following inequality
 $
 \rC^{min}_{\rm SD} \geq 2\left(2^K-1 \right),  
 $  
  at low SNRs and the equality $\rC^{min}_{\rm SD} = K$ at high SNRs.
  
  {\it Proof:} In order to calculate $\rC^{min}_{\rm SD}$ of a lattice-dependent based SD method, at a low SNR regime, we have to find 
  $
  \rP\left(\bs_{\rm ML}^k|\mathcal{I}_{\rm ld}\right), 
  $
  where $\mathcal{I}_{\rm ld}= \left\{ \bH , R_{\rB}\right\}$ and $R_{\rB} =\|\by-\bH\bs_{\rB}\|^2$. Note that at zero SNR, we have $\by=\bn$ and, consequently, $R_{\rB}=\|\bn\|^2$. Therefore, we have 
  \beq
  \rP\left(\bs_{\rm ML}^k|\mathcal{I}_{\rm ld}\right) =\int_{\Gamma_{\bs_{\rm ML}^k}}f\left(\by\left|\bH,\|\bn\|=\nu\right.\right)d\by,
  \eeq
  where $\Gamma_{\bs_{\rm ML}^k}$ is the ML decision region, that is Voronoi region, for the symbol $\bs_{\rm  ML}^k$. Using the independence of $\bH$ and $\bn$, we have 
  \beq \label{intset}
  \lim \limits_{\rho\rightarrow 0}\rP\left(\bs_{\rm ML}^k|\mathcal{I}_{\rm ld}\right) =\int_{\Gamma_{\bs_{\rm ML}^k}}f\left(\bn\left|\|\bn\|=\nu\right.\right)d\bn.
  \eeq
  
  In \cite{surface}, it is shown that $f\left(\bn\left|\|\bn\|=\nu\right.\right)=\frac{1}{S_K}\delta\left(\|\bn\|-\nu\right)$ where $\delta(\cdot)$ is the Dirac delta function and $S_K= \frac{2\pi^{\frac{K}{2}}\nu^{K-1}}{\Gamma\left(\frac{K}{2}\right)}$. For an $M$-ary modulation scheme, the integral (\ref{intset}) should be calculated over $M^k$ Voronoi regions corresponding to each $\bs_{\rm ML}^k$. In the following, we show that the integral (\ref{intset}) over the Voronoi regions corresponding to $\bs_{\rm ML}^k$ and its rotated version $\bar{\bs}_{\rm ML}^k= \bT \bs_{\rm ML}^k$ both lead to the same result, where the orthogonal matrix $\bT$ is the rotation matrix.
  
   We have
  \beq
  \int_{\Gamma_{\bs_{\rm ML}^k}}f\left(\bn\left|\|\bn\|=\nu\right.\right)d\bn =\int_{\Gamma_{\bs_{\rm ML}^k}}\frac{1}{S_K}\delta\left(\|\bn\|-\nu\right)d\bn.
  \eeq 
  We can change the variable as $\bn = \bT^H\bar{\bn}$ to get
  \beq
  \int_{\Gamma_{\bar{\bs}_{\rm ML}^k}}\frac{1}{S_K}\delta\left(\|\bar{\bn}\|-\nu\right)  d\bn =\int_{\Gamma_{\bar{\bs}_{\rm ML}^k}}\frac{1}{S_K}\delta\left(\|\bn\|-\nu\right)d\bn.
  \eeq
  The above equality holds since $\|\bn\|=\|\bar{\bn}\|$. Therefore,
  \beq \label{newAA}
  \lim \limits_{\rho\rightarrow 0}\rP\left(\bs_{\rm ML}^k|\mathcal{I}_{\rm ld}\right)=
  \lim \limits_{\rho\rightarrow 0}\rP\left(\bar{\bs}_{\rm ML}^k|\mathcal{I}_{\rm ld}\right)
  \eeq
  
  According to the symmetrical properties of the constellation, each $\bs_{\rm ML}^k$ in each hyperoctant\footnote{A $k$-dimensional space is divided into $2^k$ hyperoctants or orthants. In geometry, a hyperoctant in $k$-dimensional Euclidean space is the analog  of a half axis in one dimension, a quadrant in two dimensions or an octant in three dimensions, and the first hyperoctant is analogous to the first quadrant in two dimensions.} is a rotated version of an $\bs_{\rm ML}^k$ in the first hyperoctant. Thus, according to (\ref{newAA}), it is sufficient to solely  calculate the integral ($\ref{intset}$) over the Veronoi regions in the first hyperoctant.  
    For the binary modulation case, we denote the Voronoi region corresponding to the  point ${\mathbf 1}^k=[1,\dots,1]^\rT$ by $\Gamma_{{\mathbf 1}^k}$. Note that $\Gamma_{{\mathbf 1}^k}$ is the first hyperoctant. Since for all $\bs_{\rm ML}^k$s in the first hyperoctant we have $\Gamma_{{\bs}^k}\in \Gamma_{{\mathbf 1}^k}$, we get
  \beq
  \int_{\Gamma_{\bs_{\rm ML}^k}}f\left(\bn\left|\|\bn\|=\nu\right.\right)d\bn
  \leq \int_{\Gamma_{{\mathbf 1}^k}}  f\left(\bn\left|\|\bn\|=\nu\right.\right)d\bn.
  \eeq
   
   Therefore, in order to calculate (\ref{intset}), we can change the variable as $d\bn = \br^{k-1}d\br d\boldsymbol{\Omega}$ to obtain
  \beq
 \lim \limits_{\rho\rightarrow 0} \rP\left(\bs_{\rm ML}^k|\mathcal{I}_{\rm ld}\right) \leq \frac{1}{S_k}\int_0^\infty \int_{\boldsymbol{\Omega}_{\mathbf 1}^k} \br^{k-1}\delta(\br -\nu)d\br d{\boldsymbol{\Omega}}.
  \eeq
  Since, $S_K$ is the volume of  a $K$-dimensional hypersphere and  for a $k$ dimensional lattice with binary entries, the $2^k$ possible Voronoi regions have equal volumes,  we obtain
  \beq
  \int_0^\infty \int_{\boldsymbol{\Omega}_{\mathbf 1}^k} \br^{k-1}\delta(\br -\nu)d\br d{\boldsymbol{\Omega}} =  \frac{S_K}{2^k}.
  \eeq
 
  Consequently, we have
  $
  \lim \limits_{\rho\rightarrow 0}\rP\left(\bs_{\rm ML}^k|\mathcal{I}_{\rm ld}\right) \leq \frac{1}{2^k},
  $
  which results in 
  \beq
  \lim \limits_{\rho \rightarrow 0}-\rE_{\bs^k_{\rm ML}}\left\{ \log_M\rP\left(\bs_{\rm ML}^k|\mathcal{I}_{\rm ld}\right)\right\} \geq k \log_M 2.
  \eeq
 
  Therefore,  the MAC of a lattice-dependent radius based SD algorithm is lower bounded by
  \beq
  \lim \limits_{\rho \rightarrow 0} \rC^{min}_{\rm SD}
  \geq \sum_{k=1}^K  2^k =2(2^K-1). \eeq
    
    Now, we focus on a lattice-independent radius based   SD method. The SD presented in \cite{hasibi}  considers the FP algorithm, with a proposed radius that tends to infinity as $\rho\rightarrow 0$. Therefore, the radius does not contain any information about $\bs_{\rm ML}$ at low SNRs, or equivalently, $\mathcal{I}_{\rm li}=\left\{\bH\right\}$ and
  \beq 
  \rP\left(\bs_{\rm ML}^k|\mathcal{I}_{\rm li}\right) =\int_{\Gamma_{\bs_{\rm ML}^k}}f\left(\by\left|\bH\right.\right)d\by.
  \eeq  
   
   When $\rho\rightarrow 0$, we have $\by = \bn$ and using the independence of $\bn$ and $\bH$ we have 
  \begin{align} \label{rabete17} 
  \lim \limits_{\rho\rightarrow 0}\rP\left(\bs_{\rm ML}^k|\mathcal{I}_{\rm li}\right) &=\lim \limits_{\rho \rightarrow 0}\int_{\Gamma_{\bs_{\rm ML}^k}}f\left(\bn\right)d\bn \nonumber\\
  &= \lim \limits_{\rho \rightarrow 0} \prod_{i=1}^k
  \int_{\Gamma_{\bs_{\rm ML}^1}}f\left(n_i\right)dn_i.
  \end{align}
  
  Therefore, since the volume of the Voronoi regions corresponding to inner lattices tends to zero at low SNRs, according to (\ref{rabete17}), we obtain 
  \beq
  \lim \limits_{\rho \rightarrow 0}\rP\left(\bs_{\rm ML}^k|\mathcal{I}_{\rm li}\right)= \left\{
  \begin{array}{ll}
  	0&\bs_{\rm ML}^1 \in {\rm inner}\\
  	\left(\frac12\right)^k&\bs_{\rm ML}^1 \in {\rm outer},
  \end{array}
  \right. 
  \eeq   
  where the  inner and outer sets, correspondingly, are  the inner and outer modulation points in the  constellation set.  Consequently,  
  \begin{align}
  &\phantom{=}\;\;\lim \limits_{\rho \rightarrow 0}-\rE_{\bs^k_{\rm ML}}\left\{ \log_M\rP\left(\bs_{\rm ML}^k|\mathcal{I}_{\rm li}\right)\right\}\nonumber \\
   &= -\sum_{\bs_{\rm ML}^k}\rP\left(\bs_{\rm ML}^k|\mathcal{I}_{\rm li}\right) \log_M \rP\left(\bs_{\rm ML}^k|\mathcal{I}_{\rm li}\right) =k\log_M 2, 
  \end{align}
 which yields
    \beq
  \lim \limits_{\rho \rightarrow 0} \rC^{min}_{\rm SD}
  = \sum_{k=1}^K  2^k =2(2^K-1). \eeq
  
  Consider now the high SNR regime.  In a lattice-dependent radius based SD method, the initial radius is $R_\rB = \|\by -\bH\bs\|^2$, and  when the SNR tends to infinity,  $\bs_{\rB} =\bs_{\rm ML}=\bs$, that means that only one point, merely ${\bs}_{\rm ML}$, exists in the sphere, with a probability one. In other words, for the  given information set, i.e., $\mathcal{I}_{\rm ld}= \left\{R_\rB,\bH\right\}$, no uncertainty remains on $\bs_{\rm ML}$ when the SNR tends to infinity. Therefore, we have $ -\rE_{\bs^k_{\rm ML}}\left\{ \log\rP\left(\bs_{\rm ML}^k|R_\rB\right)\right\}=0$ and $\lim \limits_{\rho\rightarrow \infty}\rC^{min}_{\rm SD} = K$. On the other hand,  the proposed lattice  independent radius in \cite{hasibi} is $R_{\rm li} = \frac{\alpha K}{\rho}$, where $\alpha$
  is determined such that we find a lattice point inside the sphere with probability $1-\epsilon$ \cite{hasibi}, i.e., 
  \beq
  F_{\chi_K^2}\left(\frac{K\alpha}{2}\right)
  = \int_0^{K\alpha} \frac{\lambda^{\frac{n}{2}-1}}{\Gamma(\frac{n}{2})}e^{-\lambda} d\lambda=1-\epsilon, \eeq 
  where $F_{\chi_K^2}(\cdot)$ is the cumulative distribution function (CDF) of a chi-squared random variable and $\Gamma(\cdot)$ is the gamma function. For instance, if $\alpha=\sqrt{\rho}$,  at high SNRs, $\epsilon$ tends to zero and at the same time $R_{\rm li}$ approaches zero. Therefore, with probability one, only one lattice is inside the sphere at high SNRs which is $\bs_{\rm ML}$. Hence,  having the  information set $\mathcal{I}$ which is $d_{\rm li}$ at high SNRs does not leave any uncertainty on  $\bs_{\rm ML}$ and, therefore, $ -\rE_{\bs^k_{\rm ML}}\left\{ \log\rP\left(\bs_{\rm ML}^k|d_{\rm li}\right)\right\}=0$. $\blacksquare$

Theorem 1 claims that for  the lattice-dependent and independent based variants of  the SD algorithm, such as the FP algorithm in \cite{hasibi}  and the SE refinement in \cite{Damen}, we have  $\rC^{min}_{\rm SD}=K$ when the SNR tends to infinity. The complexity analysis of the SD in \cite{hasibi} at high SNRs, for the FP algorithm shows that the number of visited nodes of the FP algorithm with the proposed radius tends to $K$ at high SNRs. This means that these methods are effective in the sense that they  achieve the minimum possible complexity at high SNRs with the information set $\mathcal{I}=\{\bH,R_{\rm li}\}$ fed to the SD algorithm. The main weakness of these methods shows, indeed, at the low SNRs.  The weakness of the classic FP and SE  algorithms is because of the information fed to  these methods which leads to  an exponential complexity with respect to the problem size.   Analyses of the complexity of the FP based  SD algorithm in \cite{jalden} also shows that this method suffers from a high computational complexity at low SNRs.  This theory shows that even if   the best body search strategy, preprocessing or ordering techniques are applied to the SD algorithm, for the given lattice-independent and dependent radii, an exponential computational complexity  occurs at low SNRs.

{\it Remark 1:} Clearly, for an SD algorithm which searches over the lattice points inside a sphere of radius $R$, the {\it best}, yet infeasible, radius which contains merely one  lattice point  is $R_{\rm ML}= \|\by - \bH {\bs}_{\rm ML}\|^2$.  The proof of Theorem 1  conveys that, even with  $R_{\rm ML}$, the idea of searching inside a sphere  leads to an exponential computational complexity at low SNRs. In other words,  the  initial radius $R$ is ineffective at low SNRs.

 A fundamental question arises here: is it possible to unravel this exponential behavior at low SNRs without any performance loss? It is known that, unlike the ZF, the MMSE method declares the ML solution at low SNRs, i.e., ${\bs}_{\rm MMSE} = {\bs}_{\rm ML} $, when $\rho \rightarrow 0$ \cite{tsebook}. Therefore, there is a prospect of the existence of a method which finds the ML solution at low SNRs without visiting any nodes but just by,  intelligently, relying on ${\bs}_{\rm MMSE}$.  Theorem 1 claims that, regardless of the fact that ${\bs}_{\rm MMSE} ={\bs}_{\rm ML}$ at low SNRs, deploying the MMSE detected symbol only as the initial radius will not be successful as far as  the computational complexity is concerned. In other words, a lattice-dependent based SD algorithm is not able to benefit from the information provided by the MMSE algorithm perfectly. Indeed, the initial point obtained by the MMSE method, i.e., $\bs_{\rm MMSE}$, has more information to be exploited.
 
  Presumably,  adding the MMSE detected symbol  to the information set $\mathcal{I}$, not only in $R_\rB$ but as a separate initial point, may able to compensate for this exponential behavior of $\rC^{\min}_{\rm SD}$ at low SNRs.  Along with the  evidence we have for the existence of  a {\it lossless} method which relies on the ${\bs}_{\rm MMSE}$ and, consequently,  detects the ${\bs}_{\rm ML}$ without visiting any nodes, there is also an expectation of existence of an algorithm which is  capable of  relying on some or all of the detected symbols of the MMSE method  at all SNRs which reduces the overall effective problem size without any performance loss. A method which is capable of relying on some or all of the symbols of the initial point without any performance loss is  referred to as a lossless size reduction (LSR) aided method.  The concept of an LSR-aided SD is addressed in the next section.
  \vspace*{-4mm}
\section{MAC in a LSR-aided SD } \label{llsrsection}
 The information set of the lattice-independent and dependent algorithms can be enriched by adding a separate initial point to  set $\mathcal{I}$.  In addition, having this point in the initial information set can be interpreted  as a separate  processing block  in the preprocessing stage which enables obtaining the ML solution without visiting any nodes. The combination of a decoder that  comprises  this block and an SD algorithm is referred to as an LSR-aided SD algorithm. By partially, or completely, relying on the initial point, e.g., ${\bs}_{\rm MMSE}$, this separate block makes a decision for some, or all, of the transmitted symbols. These symbols are already declared to be detected and do not play any role in the SD search; thus, the problem size is changed. Taking into account the LSR in an SD algorithm,  the problem size of the SD algorithm, $K$, becomes a random variable. This random problem size depends on the number of symbols of the initial point on which the LSR-aided algorithm relies. 
 
 Assuming that the effective problem size after applying the LSR reduces to $K_r$, the MAC  for a given $K_r$ and $\mathcal{I}_{\rm LSR}$ is obtained by
 \beq \label{car2}
  \rC^{min}_{\rm SD}(K_r\left| \mathcal{I}_{\rm LSR}\right.) =  \sum_{k=1}^{K_r}M^{ -\rE_{\bs^k_{\rm ML}}\left\{ \log\rP\left(\bs_{\rm ML}^k|\mathcal{I}_{\rm LSR}\right)\right\}},
  \eeq
  where $\mathcal{I}_{\rm LSR}= \left\{\bH,{\bs}_{\rm B}\right\}$. Therefore,  the MAC for an LSR-aided SD is  
    \beq \label{BB}
      \rC^{min}_{\rm  LSR} = \rE_{\mathcal{I}_{\rm LSR}} \left\{\sum_{k=1}^K\rC^{min}_{\rm SD}(k\left|\mathcal{I}_{\rm LSR}\right.)\rP(K_r=k\left|\mathcal{I}_{\rm LSR}\right.)\right\}.
      \eeq
  
  The event when $K_r=k$ can be interpreted as relying on $K-k$ of the initial point.  For example $\rP(K_r=0)=1$, at a given SNR, implies that we have trusted the initial point entirely, with probability one at that SNR, and consequently, $\rC^{min}_{\rm SD}=0$. Along with considering the optimal error performance, the following definition takes into account the complexity to provide a more general view of the concept of optimality in the class of ML performance achieving algorithms.  
  
  {\it Definition 1:} An  initial information set is said to obey the {\it zero-complexity achievable} (ZCA) property at a given SNR, if  at that SNR $\rC^{min}_{\rm LSR}=0$, and an SD algorithm which exploits the ZCA property at zero and infinity SNRs, is considered to have the marginal complexity optimality, i.e., achieves the ML performance without visiting any nodes at $\rho\rightarrow 0$ and $\rho \rightarrow \infty$.
  
  Based on Definition 1, the following lemma analyzes the capability of an information set containing the MMSE or ZF points in terms of ZCA. 
  
  {\it Lemma 2:} The initial information set corresponding to an MMSE-based LSR algorithm, i.e., $\mathcal{I}_{\rm LSR}= \left\{\bs_{\rm MMSE},\bH \right\}$, is ZCA at $\rho =0$ and $\rho \rightarrow \infty $ and the information set of  a ZF-based LSR, i.e., $\mathcal{I}_{\rm LSR}= \left\{\bs_{\rm MMSE},\bH \right\}$, is only guaranteed to be ZCA at $\rho \rightarrow \infty $.
  
  {\it Proof:} See Appendix \ref{lemma2proof}.
  
  Analyzing the MAC, we have been looking for a low computational complexity point to be added to the information set to compensate for the low SNR weakness. This lemma shows that if we add ${\bs}_{\rm MMSE}$ to the information set, i.e., $\mathcal{I} = \left\{{\bs}_{\rm MMSE},\bH \right\}$, the inherent deficiency of the conventional SD algorithms at low SNRs is resolved.
   These results are summarized in Table \ref{table1},  
  \begin{table}[h]
  \caption{comparison between different information sets }
  \begin{center}
  \begin{tabular}{|c|c|c|}
  \hline
  Initial information set & Low SNR & High SNR\\ \hline
  $\mathcal{I}_{\rm LSR} =\left\{\bs_{\rm ZF},\bH \right\}$     &  ---  &  ZCA    \\ \hline
  $\mathcal{I}_{\rm LSR} =\left\{\bs_{\rm MMSE},\bH \right\}$     &  ZCA  &  ZCA    \\ \hline
  $\mathcal{I}_{\rm li} =\left\{R_{\rm li},\bH \right\}$     &  ---  &  ---    \\ \hline
  $\mathcal{I}_{\rm ld} =\left\{R_{\rm ld},\bH \right\}$      &  ---  &  ---    \\ \hline
  \end{tabular}
  \label{table1}
  \end{center}
  \end{table}
and it can be seen that the conventional SD algorithms with  lattice-dependent or independent radii, are not able to achieve the ML performance without visiting any nodes at zero and infinity SNRs. On the other hand, an algorithm which is fed with $\mathcal{I}=\left\{\bs_{\rm MMSE},\bH \right\}$, i.e., MMSE based LSR-aided SD, is capable of achieving the ML performance without visiting any nodes at zero and infinity SNRs and the ZCA property is only guaranteed for the ZF point at $\rho\rightarrow \infty$.  
\vspace{-4mm}
  \section{A Lossless Size Reduction-aided SD} \label{mcosd}
  So far, we have described the capabilities of an LSR-aided SD in terms of MAC. More precisely, we have evaluated the effect of adding the initial point ${\bs}_{\rm LD}$ to the information set $\mathcal{I}$ in Lemma 2. In the detection process, adding this information, can be translated to adopting a separate preprocessing step to determine the status of the symbols of the initial lattice point beforehand. This notion has been  considered  in some existing works, as an approach to reduce the computational complexity, e.g., \cite{rd_mls}, \cite{2012}, and \cite{neinavaie}. However, this point of view usually arrives at a sub-optimum, near ML, error performance. By focusing on the LSR property, our aim is to derive a detector which partially or entirely relies on the initial lattice point without a performance loss. From the LSR perspective,  achieving the ML performance with a reduced problem size and, even sometimes without visiting any nodes, is feasible.  In this section, we propose an LSR-aided SD algorithm. 
  
     In the detection problem (\ref{problem}), the equalized vector $\tilde{\by}$ can be written as
  $
  \tilde{\by} = \bW_p \by,
  $
   where $\bW_p$ is the preprocessing equalization matrix. The joint detection problem can be decoupled into a single detection problem which often has a lower performance compared to the ML detector. The LSR-aided SD either relies on  the detected symbols of the initial point or preserves the symbol in  a set, hereafter denoted by $\mathcal{S}$, to be searched by the SD algorithm.  
   
   The $k$th element of $\tilde{\by}$, is 
   \beq
  [\tilde{\by}]_k = [\bs]_k +[\tilde{\bn}]_k,
   \eeq
   where $k=1,\dots,K$. Assuming that $\rE\{|[\bs]_k|^2\}=1$, the $k$th received SNR  depends on the  underlying LD method. The proposed  algorithm uses the instantaneous SNR  to assess  the reliability of each symbol.   For instance,  when the ZF equalizer is adopted, then   $ {\rm SNR}_k = \frac{\rho}{[(\bH\bH)^{-1}]_{kk}}$, and for the MMSE equalizer, we have ${\rm SNR}_k=\frac{\rho}{[(\bH\bH+\frac{1}{\rho}\bI)^{-1}]_{kk}}-1$, where $\rho = \frac{1}{\sigma^2}$ \cite{zeroforcing}. 
   
   In order to perform a lossless size reduction, we need to find the reliable symbols of the initial point. The indices of the reliable detected symbols form a set denoted by $\mathcal{G}$. Indeed, the proposed detection method relies on the detected symbols whose indices are in $\mathcal{G}$, and the rest of the symbols construct the set $\bar{\mathcal{G}}$. In Theorem 2, we determine $\mathcal{G}$  such that the proposed method guaranties the ML performance. The set $\mathcal{G}$ is constructed as 
    \beq \label{setG}
    \mathcal{G} \triangleq \left\{k\left| {\rm SNR}_k> \eta_k ,k=1,\dots,K \right.\right\}, 
    \eeq 
    where $\eta_k$ is obtained from the following equation
    \beq \label{etaequation1}
    \rP\left( \mathcal{E}^k_{\rm LD}, {\rm SNR}_k >\eta_k \right) = 
    \rP\left( \mathcal{E}^k_{\rm ML}, {\rm SNR}_k >\eta_k \right),
    \eeq
    where $\mathcal{E}^k_{\rm LD}$ and $\mathcal{E}^k_{\rm ML}$ are the $k$th symbol error events of the LD and ML detection methods, respectively.
    
     It should be noted that finding $\eta_k$ from (\ref{etaequation1}) is an offline process. Indeed, using (\ref{etaequation1}), a lookup table can be provided to find $\eta_k$ at each SNR. More details about finding $\eta_k$ will be discussed later.   
    The reduced search space is $\mathcal{S} =\prod_{i=1}^K  {D}_i $ which is $K$ary Cartesian product over $K$ set $\{{D}_i\}_{i=1}^K$, with
   \beq \label{Dset}
   {D}_i = \left\{\begin{array}{lr}
   [\bs_{\rm LD}]_i&i\in \mathcal{G}\\
   \mathcal{C}&i\notin \mathcal{G}
   \end{array}	\right. ,
   \eeq
    where $[\bs_{\rm LD}]_i  = \arg \min \limits_{s\in \mathcal{C}}|[\tilde{\by}]_i-s|$ denotes the $i$th symbol of the initial point $\bs_{\rm LD}$. 
    
    The following theorem  provides the sufficient reduced search space which relies on some or all of the initial point symbols and leads to achieving the exact ML performance. 
    
     {\it Theorem 2:} An algorithm which relies on the initial point symbols whose indices are in $\mathcal{G}$, and searches over the remaining symbols in the set $\mathcal{S}$ , where $|\mathcal{S}|=M^{K-|\mathcal{G}|}$, achieves the optimal ML performance. 
    
  {\it Proof:} 
  In order for the algorithm to achieve the ML performance, $\eta_k$ should be selected such that $\rP\left( [\hat{\bs}]_k \neq [\bs]_k \right)=\rP\left( [{\bs_{\rm ML}}]_k \neq [\bs]_k \right)$ for all $k$. The $k$th symbol error probability of the proposed method can be expanded as
  \begin{align} \label{1exp}
  \rP\left( [\hat{\bs}]_k \neq [\bs]_k \right) &= \rP\left( [\hat{\bs}]_k \neq [\bs]_k, [{\bs}]_k \in D_k  \right) \nonumber \\
   &\phantom{=}+ \rP\left( [\hat{\bs}]_k \neq [\bs]_k, [{\bs}]_k \notin D_k  \right), 
  \end{align} 
where $D_k$ is defined in (\ref{Dset}), $\bs$ and $\hat{\bs}$ are the transmitted symbol and the detected symbol by the proposed algorithm, respectively. The first term of the right hand side of (\ref{1exp}) can also be expanded as
\begin{align} \label{2exp}
&\phantom{=}\;\;\rP\left( [\hat{\bs}]_k \neq [\bs]_k, [{\bs}]_k \in D_k  \right)\nonumber \\
 &= \rP\left( [\hat{\bs}]_k \neq [\bs]_k, [{\bs}]_k \in D_k, |D_k|=1 \right)\nonumber \\ &\phantom{=}+\rP\left( [\hat{\bs}]_k \neq [\bs]_k, [{\bs}]_k \in D_k, |D_k| = M  \right). 
\end{align}

According to the proposed detection method, in the first term on the right hand side of (\ref{2exp}), the event $\left\{|D_k|=1\right\}$ means that $D_k = \left\{[\bs_{\rm LD}]_k \right\}$, where $[\bs_{\rm LD}]$ is the detected symbol of the LD algorithm. Consequently, the joint event $\left\{ [\bs]_k \in D_k,|D_k|=1  \right\}$ implies that $[\hat{\bs}]_k=[{\bs}_{\rm LD}]_k = [\bs]_k$ which results in  
\beq
\rP\left( [\hat{\bs}]_k \neq [\bs]_k, [{\bs}]_k \in D_k, |D_k|=1 \right) = 0.
\eeq

Now, we focus on the second term on the right hand side of (\ref{2exp}). It should be noted that the event  $\left\{ |D_k|=M \right\}$ means that all the modulation points exist in the set $D_k$ and, therefore, the event $\left\{[\bs]_k \in D_k\right\}$ occurs. Hence,  $\left\{[\bs]_k \in D_k,|D_k|=M\right\}=\left\{|D_k|=M\right\}$ and
\begin{align}
&\rP\left([\hat{\bs}]_k \neq [\bs]_k, [\bs]_k \in D_k,|D_k|=M \right) \nonumber \\
 &= \rP\left([\hat{\bs}]_k \neq [\bs]_k, |D_k|=M \right). 
\end{align}
Note that the proposed algorithm performs the ML search over the set $\mathcal{S}$. The error event in the detection of the $k$th symbol occurs when all the search candidates, in $\mathcal{S}$, sharing the same $k$th symbol with the transmitted vector, do not meet the ML minimum distance detection criteria. We denote the set of all vectors sharing the same $k$th symbol with $\bs$ by $\Lambda_{[\bs]_k}$, that is, 
$
\Lambda_{[\bs]_k} = \left\{\bx \in \mathcal{C}^K\left| [\bx]_k = [\bs]_k \right. \right\}.
$
 Therefore, we have
 \begin{align} \label{37exp}
  &\phantom{=}\;\rP\left([\hat{\bs}]_k \neq [\bs]_k, |D_k|=M \right) \nonumber \\
  &= \rP\left(\bigcap\limits_{\bs \in \Lambda_{[\bs]_k} }\bigcup\limits_{\bs_i \in \mathcal{S} } \|\by -\bH \bs_i\| \leq \|\by -\bH \bs\| ,|D_k|=M \right).
 \end{align}  
  It should be noted that we have $\mathcal{S} \subseteq \mathcal{C}^K$, therefore
  \begin{align} \label{38exp}
  &\phantom{=}\;\rP\left(\bigcap\limits_{\bs \in \Lambda_{[\bs]_k} }\bigcup\limits_{\bs_i \in \mathcal{S} } \|\by -\bH \bs_i\| \leq \|\by -\bH \bs\| ,|D_k|=M \right) \nonumber \\
   &\leq\rP\left(\bigcap\limits_{\bs \in \Lambda_{[\bs]_k} }\bigcup\limits_{\bs_i \in \mathcal{C}^K } \|\by -\bH \bs_i\| \leq \|\by -\bH \bs\| ,|D_k|=M \right)
  \nonumber \\
  &= \rP([\bs_{\rm ML}]_k \neq [\bs]_k, |D_k|=M ).
  \end{align}
  
  Next, we focus on the second term on the right hand side  of (\ref{1exp}). We can write
  \begin{align}
  &\phantom{=}\;\;\rP\left( [\hat{\bs}]_k \neq [\bs]_k, [{\bs}]_k \notin D_k  \right) \nonumber \\
  &= \rP\left( [\hat{\bs}]_k \neq [\bs]_k, [{\bs}]_k \notin D_k, |D_k|=1  \right) \nonumber \\
  &\phantom{=}\;+\rP\left( [\hat{\bs}]_k \neq [\bs]_k, [{\bs}]_k \notin D_k, |D_k|=M  \right).
  \end{align}
  According to the proposed method, joint event $\left\{[{\bs}]_k \notin D_k, |D_k|=M \right\}$ does not happen, therefore,
   \begin{align} \label{40exp}
 &\phantom{=}\;\;\rP\left( [\hat{\bs}]_k \neq [\bs]_k, [{\bs}]_k \notin D_k  \right) \nonumber \\
 & = \rP\left( [{\bs}]_k \notin D_k , |D_k|=1 \right) = \rP\left( [{\bs}_{\rm LD}]_k \neq [\bs]_k, {\rm SNR}_k>\eta_k  \right). 
 \end{align}
 Since, the event $\{ |D_k| =M\}$ is equivalent to the event ${\rm SNR_k}\leq \eta_k$, it follows from (\ref{1exp}), (\ref{37exp}), (\ref{38exp}), and (\ref{40exp}) that
 \begin{align}
 \rP\left( [\hat{\bs}]_k \neq [\bs]_k\right)&\leq  \rP\left( [{\bs}_{\rm LD}]_k \neq [\bs]_k, {\rm SNR}_k>\eta_k  \right)  \nonumber \\
 &\phantom{=}\;\;+\rP\left( [{\bs}_{\rm ML}]_k = [\bs]_k, {\rm SNR}_k<\eta_k  \right). 
 \end{align}
 Therefore, in order to achieve the ML performance it is sufficient  to have
 \begin{align} \label{42exp}
 &\rP\left( [{\bs}_{\rm LD}]_k \neq [\bs]_k, {\rm SNR}_k>\eta_k  \right) \nonumber \\& +\rP\left( [{\bs}_{\rm ML}]_k \neq [\bs]_k, {\rm SNR}_k<\eta_k  \right) 
  \leq \rP([\bs_{\rm ML}]_k\neq[\bs]_k ). 
 \end{align}
 
 Since, the $k$th symbol error probability of the ML detector can be expanded as 
 \begin{align}
  \rP([\bs_{\rm ML}]_k\neq[\bs]_k )& =\rP([\bs_{\rm ML}]_k\neq[\bs]_k, {\rm SNR}_k >\eta_k) \nonumber \\
  &\phantom{=}\;\;+\rP([\bs_{\rm ML}]_k\neq[\bs]_k, {\rm SNR}_k <\eta_k),
  \end{align} 
 by defining the events $\mathcal{E}^k_{\rm LD} = \left\{[\bs_{\rm LD}]_k \neq [\bs]_k\right\}$ and $\mathcal{E}^k_{\rm ML} = \left\{[\bs_{\rm ML}]_k \neq [\bs]_k\right\}$, which are the symbol error events corresponding to the LD and the ML methods, respectively, the sufficient condition  for the proposed method to achieve the ML performance is obtained by (\ref{etaequation1}).
 
   We have shown that a threshold which is obtained by (\ref{etaequation1}), guarantees the ML performance. However, in order to prove that the algorithm has the lossless size reduction property, we have to show that $\hat{\bs}=\bs_{\rm ML}$. It should be noted that the ML decision has the uniqueness property \cite{anderson}. In other words, an algorithm which achieves the same performance as the ML detector, has also the same decision with probability one, i.e. $\rP(\hat{\bs}=\bs_{\rm ML})=1$.  
  $\blacksquare$
  
  Theorem 2 proposes a {\it fading-type and application independent} algorithm. In other words, the derived threshold is general in the sense that  it is applicable  to different fading types, e.g., Rayleigh,  Nakagami, etc, and  applications, e.g., coded and uncoded MIMO, OFDM systems, etc. 
    The detection algorithm is summarized in Algorithm 1.    
    \begin{algorithm} \label{algorithm}
    	\caption{LSR-aided SD}
    	{\textbf{Input:}} The initial point $\bs_{\rm LD}$; the channel matrix $\bH$.
    	\\{\textbf{Output:}} The detected symbol $\bs$.\;
    	\\ $\bullet$ \textit{Initialization}\;
    	\\  Determine the sets $\mathcal{G}=\left\{g_1,\dots,g_{|\mathcal{G}|}\right\}$ and $\bar{\mathcal{G}}=\left\{\bar{g}_1,\dots,\bar{g}_{|\bar{\mathcal{G}}|}\right\}$.\;
    	\\ $\bullet$ \textit{Lossless Size Reducton}\;
    	\\  Select the valid symbols of $\bs_{\rm LD}$ to form the vector: $\bs_{\rm v} = \left[ s_{{\rm LD}_{g_1}},\dots,s_{{\rm LD}_{g_{|\mathcal{G}|}}} \right]$.  Use the columns of $\bH$ whose indices belong to $\mathcal{G}$ to get  $\bH_\rv =[\bh_{g_1},\dots,\bh_{g_{|\mathcal{G}|}}] $ and strike these columns out to get $\bH_\br$.\;
    	\\ $\bullet$ \textit{Sphere Decoder}\;
    	\\  Apply an SD algorithm to solve $\hat{\bs}_\rr=\arg\min \limits_{\bs_\rr}\|\by_\rr - \bH_\rr \bs_\rr\|^2$ where $\by_\rr =\by - \bH_\rv \bs_\rv$. Combine $\hat{\bs}_\rv$ and $\hat{\bs}_{\rr}$ to get $\hat{\bs}$.
    \end{algorithm}
    
    It should be noted that regardless of which SD algorithm we use, the lossless size reduction step reduces the overall computational complexity.
  
   As it was mentioned previously finding $\eta_k$ from (\ref{etaequation1}) can be accomplished via a lookup table which is provided off line by  Monte Carlo simulations. One way to build the lookup table at a given SNR is to compute the probabilities $\rP\left(\mathcal{E}_{\rm LD}^k,{\rm SNR}_k>\eta_k\right)$ and $\rP\left(\mathcal{E}_{\rm ML}^k,{\rm SNR}_k>\eta_k\right)$ for different values of $\eta_k$, and their intersection gives the proposed $\eta_k$. In the following section we provide a more convenient way to construct the offline lookup table.
   \vspace*{-4mm}
   \subsection{Constructing the lookup table}
 In order for a lossless size reduction aided algorithm to achieve the ML performance at a given SNR, it is sufficient to choose an ${\eta_k}$, or equivalently $\frac{\eta_k}{\rho}$, which is an answer/root of (\ref{etaequation1}). The smallest root of (\ref{etaequation1}) is, hereafter, denoted by $\eta_k^\star$.  It should be noted that at a given $\rho$, (\ref{etaequation1}) may have multiple roots and an efficient algorithm should choose the smallest $\eta_k^\star$ to obtain a lower computational complexity. A trivial answer for (\ref{etaequation1}) is $\frac{\eta_k}{\rho}=\infty$ and, in general, it is not guaranteed that a finite $\frac{\eta_k}{\rho}$ exists for all values of $\rho$. Also, $\frac{\eta_k^\star}{\rho}= \infty$, at a given SNR, implies that the initial information is not capable of pruning the search tree at that SNR.
 
  In general, obtaining a closed form solution for (\ref{etaequation1}) may be complicated and depends on the fading type and the adopted LD method. In this section, we provide a general suboptimal answer which is less or equal than the smallest root of (\ref{etaequation1}). Since, $\frac{\eta_k}{\rho}=\infty$ is always a valid answer, it is guaranteed that scaling the suboptimal $\eta_k$ always arrives at a solution for (\ref{etaequation1}). Intuitively, if choose  $\frac{\eta_k}{\rho}$ larger than $\frac{\eta_k^\star}{\rho}$, it means that we have considered more restrictive conditions for relying on an LD detected symbol and, consequently, scaling a suboptimal $\frac{\eta_k}{\rho}$ leads to a better performance and a higher computational complexity. 
  
   By neglecting $\rP\left([\bs_{\rm ML}]_k\neq [\bs]_k, {\rm SNR_k}< \eta_k \right)$ in (\ref{42exp}), we can arrive at the integral 
  \beq \label{44n}
  \int_{\frac{\eta_k}{\rho}}^\infty \rP\left([\bs_{\rm LD}]_k \neq [\bs]_k |x_k\right)f(x_k) dx_k = \rP\left([\bs_{\rm ML}]\neq [\bs]_k\right).
  \eeq
  In general neglection of this term violates ML achieving property. In other words, the root of (\ref{44n}) is less or equal than $\frac{\eta_k^\star}{\rho}$.  Yet, since we are seeking a subotimal solutionl, we are not obligated to keep this term. However, later in this section, we show that this term is negligible at high SNR regime and  this approximation leads to the ML performance when $\rho \rightarrow \infty$. This implies that the root of (\ref{44n}) at $\rho \rightarrow \infty$, is $\frac{\eta_k^\star}{\rho}$.
  
  We can rewrite the integral in (\ref{44n}) as 
    \begin{align} \label{45n}
    &{\phantom{=}\;\;}\int_{\frac{\eta_k}{\rho}}^\infty \rP\left([\bs_{\rm LD}]_k \neq [\bs]_k |x_k\right)f(x_k) dx_k \nonumber \\
     &= \rP\left([\bs_{\rm LD}]\neq [\bs]_k\right) - 
    \int_0^{\frac{\eta_k}{\rho}} \rP\left([\bs_{\rm LD}]_k \neq [\bs]_k |x_k\right)f(x_k) dx_k,
    \end{align}
  where $\rP\left([\bs_{\rm LD}]\neq [\bs]_k\right) = \int_0^\infty \rP\left([\bs_{\rm LD}]_k \neq [\bs]_k |x_k\right)f(x_k) dx_k$. The term $ \rP\left([\bs_{\rm LD}]_k \neq [\bs]_k |x_k\right)$ is the error probability of a detected symbol by the LD method given $x_k$ and it is bounded as 
  \beq \label{46n}
 \rP\left([\bs_{\rm LD}]_k \neq [\bs]_k |x_k\right) \leq \left(1-\frac{1}{M} \right).
  \eeq   
   Using (\ref{46n}) in (\ref{44n}), results in an $\eta_k$ which is smaller than or equal to $\eta_k^\star$. Therefore, it is a suboptimal solution. According to (\ref{44n}), (\ref{45n}) and (\ref{46n}) we can write 
   \begin{align} 
   &\phantom{=}\;\left(1-\frac{1}{M} \right)\int_0^{\frac{\eta_k}{\rho}} f(x_k) dx_k \nonumber \\
    &= \rP\left([\bs_{\rm LD}]_k\neq [\bs]_k\right) -\rP\left([\bs_{\rm ML}]_k\neq [\bs]_k\right) 
   \end{align}
      Denoting the root of the above inequality as $\eta_k^{\rm sub}$,  we have 
      \beq \label{etasub}
      \frac{\eta_k^{\rm sub}}{\rho} =  F^{-1}_{x_k} \left(\frac{M}{M-1}\left(\rP\left(\mathcal{E}^k_{\rm LD} \right) -\rP\left(\mathcal{E}^k_{\rm ML} \right)\right) \right), 
      \eeq
      where $F_{x_k}(\frac{\eta_k}{\rho})=\int_0^{\eta_k}f(x_k)dx_k$, is the CDF of $x_k$ and it depends on the channel realization and the LD detection method. Calculation of $\frac{\eta_k^{\rm sub}}{\rho}$  only  relies  on  obtaining $F_{x_k}(\frac{\eta_k}{\rho})$, $\rP\left(\mathcal{E}^k_{\rm ML} \right)$ and $\rP\left(\mathcal{E}^k_{\rm LD} \right)$  via  Monte Carlo simulations. For the derived suboptimal threshold, we saw that $\frac{\eta_k^{\rm sub}}{\rho}\leq \frac{\eta_k^{\star}}{\rho}$ and by scaling $\frac{\eta_k^{\rm sub}}{\rho}$ we can  come arbitrary closer to the smallest root of  (\ref{etaequation1}). Hence, $\frac{\eta_k^{\rm sub}}{\rho}$ can be used in a convenient numerical method for completing the lookup table which can determine $\frac{\eta_k^{\star}}{\rho}$. 
\vspace*{-5mm}
  \subsection{Marginal Complexity Optimality of an MMSE-based LSR-aided SD}
  As it was mentioned in Section  \ref{llsrsection}, the initial information set of an MMSE based LSR-aided SD, has the ZCA property, i.e., the potential of achieving the ML performance without visiting any nodes at zero and infinity SNRs. Now, we examine the computational complexity of the proposed algorithm to see if  the algorithm is marginally  optimal, i.e., if it is capable of exploiting the ZCA property. As it was mentioned previously, if a method obeys the marginal  complexity  optimality, it does not need to tolerate any complexity more than the LD computational complexity at zero and infinity SNRs, and while achieving the ML performance, it does not need to  visit any nodes. The following theorem addresses this property for the proposed algorithm.
 
{\it Theorem 3:} The proposed MMSE based LSR-aided SD algorithm is marginally optimal.

{\it Proof:}  Assume that the effective problem size is $K_r$. If for the proposed method at a given SNR, we can show that $\rP(K_r=0)=1$, then  the number of visited nodes becomes zero with probability one at that SNR. In order to prove that the proposed method is marginally optimal, we need to show that $\rP(K_r=0)=1$ at $\rho\rightarrow 0 $ and $\rho \rightarrow \infty$. We have 
\begin{align}
\rP(K_r=0)&= \rP\left(\bigcap_{k=1}^{K}{\rm SNR}_k>\eta_k\right)\nonumber \\
 &\geq 1- \sum_{k=1}^K\rP\left({\rm SNR}_k < \eta_k \right),
\end{align} 
where the last inequality follows from the union bound. Hence, since ${\rm SNR}_k= \rho x_k$, the sufficient condition for marginal optimality is
\beq \label{infty}
\lim \limits_{\rho\rightarrow 0,\infty} \rP\left( x_k <\frac{\eta_k}{\rho}\right)=0.
\eeq

Therefore, in order to prove the marginal optimality, it is sufficient to show that for the given $\eta_k$ obtained from (\ref{etaequation1}), we have  $\lim \limits_{\rho \rightarrow 0,\infty}\frac{\eta_k}{\rho}=0$. It follows from Appendix B that the events $\mathcal{E}_{\rm ML}^k$ and $\mathcal{E}_{\rm LD}^k$ are equivalent when $\rho \rightarrow 0$. Therefore, when $\rho \rightarrow 0$, (\ref{etaequation1}) holds for all values of $\frac{\eta_k}{\rho}$, including $\frac{\eta_k}{\rho} =0$. 
 
 Now, we consider the case when $\rho \rightarrow \infty$. It should be noted that (\ref{42exp}), similar to (\ref{etaequation1}), results in an ML performance achieving $\eta_k$. At high SNR regime, for  the ML method, we have  $\rP\left([\bs_{\rm ML}]_k \neq [\bs]_k \right)\doteq \frac{1}{\rho^{ d_{\rm ML}}}$, where $d_{\rm ML}$ is the diversity gain of the ML method. We show that if $\eta_k$ satisfies 
 \beq \label{jexp}
 \lim \limits_{\rho \rightarrow \infty} \rP\left([\bs_{\rm LD}]\neq [\bs]_k, {\rm SNR}_k> \eta_k \right) =\frac{1}{\rho^{d_{\rm ML}+\epsilon}},
 \eeq
 where $\epsilon$ is an arbitrary small number, then (\ref{42exp}) holds. The reason is that if (\ref{jexp}) is satisfied, the first term of (\ref{42exp}) becomes negligible at high SNR since
 \beq \label{highsnrapp}
 \lim \limits_{\rho \rightarrow \infty} \frac{\rP\left([\bs_{\rm LD}]\neq [\bs]_k, {\rm SNR}_k> \eta_k \right)}{\rP\left([\bs_{\rm ML}]\neq [\bs]_k\right)}= \lim \limits_{\rho \rightarrow \infty} \rho^{-\epsilon}=0,
	\eeq
 and, obviously, for the remaining terms at high SNR, we have $\rP\left([\bs_{\rm ML}]\neq [\bs]_k, {\rm SNR}_k> \eta_k \right)\leq \rP\left([\bs_{\rm ML}]\neq [\bs]_k\right)$. Therefore, if $\eta_k$ stisfies (\ref{jexp}), equation (\ref{42exp}) holds. 
 Hence, according to (\ref{jexp}), we have
 \begin{align}
&\phantom{=}\;\;\rP\left([\bs_{\rm LD}]\neq [\bs]_k, {\rm SNR}_k> \eta_k \right) \nonumber \\
&= \int_{\frac{\eta_k}{\rho}}^\infty \rP\left([\bs_{\rm LD}]_k \neq [\bs]_k |x_k\right)f(x_k) dx_k = \frac{1}{\rho^{d_{\rm ML}+\epsilon}}.
 \end{align}
 The conditional symbol error probability of $M$-ary QAM is \cite{gallager}
 \begin{align} \label{lemma31}
 \rP([\bs_{\rm LD}]_k\neq [\bs]_k |x_k ) &= 2\beta\mathcal{Q}\left(\sqrt{\alpha \rho x_k} \right) - \beta^2\mathcal{Q}^2\left(\sqrt{\alpha \rho x_k} \right) \nonumber \\
 &\leq 2\beta\mathcal{Q}\left(\sqrt{\alpha \rho x_k} \right),
 \end{align}
 where $\alpha =\frac{3}{M-1}$ and $\beta=2\left(1-\frac{1}{\sqrt{M}}\right)$. For linear equalizers, the problem turns into some single symbol detection problem, and for a conditional SNR and channel realization,  (\ref{lemma31}) holds as far as square QAM is adopted. Invoking the fact that $\mathcal{Q}(\cdot)$ is a decreasing function of its argument, the upper bound $\mathcal{Q}(\sqrt{y})\leq \frac{1}{2}\exp(-\frac{y}{2})$, holds. Moreover, since $\int_{\frac{\eta_k}{\rho}}^\infty f(x)dx\leq 1$, we obtain
\begin{align} \label{bexp}
\int_{\frac{\eta_k}{\rho}}^\infty 2\beta\mathcal{Q}\left(\sqrt{\alpha \rho x_k} \right) f(x_k)dx_k
&\leq 2\beta\mathcal{Q}\left(\sqrt{\alpha  \eta_k} \right) \nonumber \\ 
&\leq \beta \exp\left(-\frac{\alpha \eta_k}{2} \right).
\end{align}
Consequently, using (\ref{jexp}) and (\ref{bexp}) we obtain 
$
\lim \limits_{\rho \rightarrow \infty}\eta_k = \frac{2}{\alpha} \ln \left(\beta \rho^{d_{\rm ML}+\epsilon} \right),
$
and, therefore, $\lim \limits_{\rho \rightarrow \infty}\frac{\eta_k}{\rho} = 0$.
$\blacksquare$  

The proposed method is as complex as the MMSE method at zero and infinity SNRs with probability one. However, the theory does not imply that the algorithm deterministically adopts the LD method  at these SNRs, and there is always a probability that the proposed method performs as an ML search.    
\vspace*{-4mm}
\subsection{High SNR Approximation for the Threshold} \label{appsec}
 A finite $\eta_k$ that is capable of achieving the ML performance and exploiting the ZCA property can be obtained by (\ref{42exp}) or (\ref{etaequation1}). Using some approximations, we propose a threshold that can be used in a near ML method that does not need a lookup table. In the simulation section, the approximated method will be shown to have a near ML performance while achieving a very low computational complexity. As it was stated previously, according to (\ref{highsnrapp}),  the term $\rP\left([\bs_{\rm ML}]_k\neq [\bs]_k, {\rm SNR_k}< \eta_k \right)$ can be neglected in (\ref{42exp}). Therefore, we can write
\beq
\int_{\frac{\eta_k}{\rho}}^\infty \rP\left([\bs_{\rm LD}]_k \neq [\bs]_k |x_k\right)f(x_k) dx_k = \rP\left([\bs_{\rm ML}]\neq [\bs]_k\right).
\eeq
According to (\ref{lemma31}), and using similar calculation as in (\ref{bexp}), we get
$
\beta \exp\left(-\frac{\alpha \eta_k}{2} \right) =\rP\left([\bs_{\rm ML}]\neq [\bs]_k\right),
$
which yields
\beq \label{appeta}
\eta_k = \frac{2}{\alpha}\log\frac{\beta}{\rP\left([\bs_{\rm ML}]\neq [\bs]_k \right)}.
\eeq
Using (\ref{appeta}) as an approximation of the solution of (\ref{etaequation1}) will result in a suboptimal performance and a dramatically reduced computational complexity. As it was previously stated, since $\frac{\eta_k}{\rho} = \infty $ is a trivial solution of (\ref{etaequation1}). By scaling $\eta_k$, we can come arbitrary closer to the ML performance without a lookup table. In the simulation section, we show that scaled $\eta_k$ yields a performance which is very close to that of the ML detector with a significantly lower computational complexity. 
\section{simulation Results} \label{simulation}
In this section, we first compare the computational complexity bounds with the actual simulated number of visited nodes to examine the ZCA property for different methods. Next,  we verify the claimed theoretical results on the performance and the computational  complexity of the proposed method in two different channels: flat fading Rayleigh MIMO channel and frequency selective channel.    In our simulations, we  consider \cite{sphere2017} as an FP variant and the SE algorithm of \cite{Damen} as an SE variant of the SD method. The curves of complexity and symbol error probability  are plotted versus the total SNR, $\rho_T \triangleq K\rho$, and the number of transmit antennas.
\vspace*{-5mm}
\subsection{Minimum Achievable Complexity and the ZCA Property}
\begin{figure}
	\centering
	\subfloat[Lattice-independet radius]{
		\begin{tikzpicture}[scale=.43]
		\begin{axis}[axis background/.style={
			shade,top color=white!10,bottom color=white},
		legend style={fill=white},
		legend style={font=\tiny},
		legend style={at={(.35,.73)},
			anchor=south west},
		xlabel={Total SNR (dB)},
		ylabel={Average number of visited nodes},
		grid=major,
		xmin = -8, xmax = 34,
		ymin = 0,
		legend entries={ \Large MAC$\left(\mathcal{I}={\{ R_H,\bH \}}\right)$  , \Large FP Algorithm \cite{sphere2017}},
		]
		\addplot+[very thick,smooth,line join = round, color= black, mark=none]  coordinates{                                                         
			(-8 ,12.5162) (-6,11.0667)(-4,9.9155)    (-2,8.6147)    (0,7.4864)    (2,6.4630)    (4,5.6909)    (6,4.9307)    (8,4.4733)    (10,4.2563)    (12,4.0456)    (14,4.0068)    (16,4)    (18,4)    (20, 4)    (22,4) (24,4)
			(26, 4) (28, 4)(30,4)(32,4) (34,4)};
		\addplot+[very thick,mark size =3,line join = round,mark=none,color= blue,dashed]  coordinates{                                                         
			(-8,16.2021) (-6,14.8041) (-4,13.4972)    (-2,12.1784)    (0,11.0927)    (2,9.7905)    (4,8.6228)    (6,7.8283)    (8,7.1047)    (10,6.6225)    (12,6.1322)    (14, 5.7378)    (16,5.4961)    (18,5.3826)    (20, 5.2204)    (22,5.0061) (24,5.0061) (26, 5.0061) (28, 5.0061)(30,5.0061)(32,5.0061) (34,5.0061)
		}; 
		\end{axis}grid=major,
		16.2021
		14.8041
		13.4972
		12.1784
		11.0927
		9.7905
		9.1228
		7.7283
		7.4047
		6.8225
		6.1322
		5.7378
		5.4961

		5.1289
		5.1148
		5.0385
		5.0244

		\end{tikzpicture}
		\label{fincke}
	}
	\subfloat[Lattice-dependent radius]{
		\begin{tikzpicture}[scale=.43]
		\begin{axis}[axis background/.style={
			shade,top color=white!10,bottom color=white},
		legend style={fill=white},
		legend style={font=\tiny},
		legend style={at={(.35,.73)},
			anchor=south west},
		xlabel={Total SNR (dB)},
		ylabel={Average number of visited nodes},
		grid=major,
		xmin = -8, xmax = 30,
		ymin = 0,
		legend entries={\Large MAC $\left(\mathcal{I}={\{ \bs_\rB,\bH \}}\right)$  , \Large SE algorithm \cite{Damen}},
		]
		\addplot+[very thick,smooth,line join = round, color= blue, dashed, mark = none]  coordinates{                                                         
			(-8,    14.3320   )  (-6,  12.9020   )   (-4,  11.8640   )   (-2,   10.7250   )   (0,    9.7270  )    (2,   8.5300   )    (4,    7.6140   )     (6,   7.0190   )     (8,  6.2200   )    ( 10,   5.7890   )     (12, 
			5.4910   )    (14,   5.4080  )    (16, 5.3290  )    (18, 5.2430  )    (20, 5.1280  )    (22, 5.0710  )    (24,  5.0240  )    (26,   5.0030  )    (28,  5.0000 )   ( 30, 5.0  )};
		\addplot+[very thick,mark size =3,line join = round,color= black, mark=none]  coordinates{                                                         
			(-8,  11.0363  )  (-6,  9.4101  )   (-4,  8.4846  )   (-2,  7.3593  )   (0,  6.7033  )    (2,  5.8648  )    (4,   5.2131  )     (6,  4.6730  )     (8,  4.3746  )    ( 10,  4.2059  )     (12, 
			4.0572  )    (14,  4.0203 )    (16, 4 )    (18,  4 )    (20,4 )    (22, 4 )    (24, 4 )    (26, 4 )    (28,  4 )   ( 30, 4 ) 
		}; 
		\end{axis}grid=major,
		
		\end{tikzpicture}

		\label{schnorr}	
	}\\
	\subfloat[ZF-based LSR]{
		\begin{tikzpicture}[scale=.43]
		\begin{axis}[axis background/.style={
			shade,top color=white!10,bottom color=white},
		legend style={fill=white},
		legend style={font=\tiny},
		legend style={at={(.35,.73)},
			anchor=south west},
		xlabel={Total SNR (dB)},
		ylabel={Average number of visited nodes},
		grid=major,
		xmin = -8, xmax = 44,
		ymin = 0,
		legend entries={\Large MAC  $\left(\mathcal{I}={\{ \bs_{\rm ZF},\bH \}}\right)$ , \Large ZF-based LSR SD},
		]
		\addplot+[thick,smooth,line join = round, color= black, mark=none]  coordinates{                                                         
			(-8,12.6248)(-6,11.8394)(-4,11.0829)(-2,10.3550)(0,9.6559) (2,8.9855)(4,8.3439)(6,7.7310)(8,7.1468)(10,6.5914)(12,6.0648)(14,5.5668)(16,5.0976)(18,4.6572) (20,4.2455)(22,3.8625)
			(24,3.5083)(26,3.1828)(28,2.8860)(30,2.6180)(32,2.3788)(34,2.1682)(36,1.9864)(38,1.7334)(40,1.5091)(42,1.3135)(44,1.0467)};
		
		\addplot+[thick,mark size =3,line join = round,mark=none,color= blue,dashed]  coordinates{                                                         
			(-8,15.4044) (-6,14.3831) (-4,13.4332)    (-2,12.5504)    (0,11.7306)    (2,10.9697)    (4,10.2634)    (6,9.6076)    (8,8.9982)    (10,8.4311)    (12,7.9020)    (14, 7.4068)    (16,6.9414)    (18,6.5016)    (20, 6.0833)    (22, 5.6823) (24,5.2944) (26,4.9156) (28, 4.5416)(30,4.1683)(32,3.7915) (34,3.4072) 
			(36,3.0110) (38,2.5990) (40,2.1669)(42, 1.8106)(44, 1.6568)
		}; 
		\end{axis}grid=major,
		\end{tikzpicture}
		
		\label{ZFllsr}	
	}
	\subfloat[MMSE-based LSR]{
		\begin{tikzpicture}[scale=.43]
		\begin{axis}[axis background/.style={
			shade,top color=white!10,bottom color=white},
		legend style={fill=white},
		legend style={font=\tiny},
		legend style={at={((.26,.8))},
			anchor=south west},
		xlabel={Total SNR (dB)},
		ylabel={Average number of visited nodes},
		grid=major,
		xmin = -8, xmax = 44,
		ymin = 0,
		legend entries={\Large  MAC $\left(\mathcal{I}={\{ \tiny \bs_{\tiny {\rm \tiny MMSE}},\bH \}}\right)$  , \Large MMSE-based LSR-SD },
		]
		\addplot+[thick,smooth,line join = round, color= black, mark=none]  coordinates{                                                         
			(-8,0) (-6,0)(-4,0)    (-2,0.0455)    (0,7.7758)    (2,8.7187)    (4,8.6039)    (6,6.9669)    (8,6.0348)    (10,5.1435)    (12,4.7283)    (14,4.2259)    (16,3.8717)    (18,3.6412)    (20,  3.3536)    (22,  3.1409) (24,2.9073) (26,2.4882) (28,2.2818)(30,1.7950)(32,1.6776) (34,1.3112)
			(36,1.0972) (38,0.9226) (40,0.6615)(42,0.4283)(44,0.3190)};

		\addplot+[thick,mark size =3,line join = round,mark=none,color= blue,dashed]  coordinates{                                                         
			(-8,0) (-6,0) (-4,0)    (-2,0.0392)    (0,10.2568)    (2,10.3742)    (4,10.3717)    (6,10.0433)    (8,9.0883)    (10,8.1958)    (12,7.4535)    (14, 6.8017)    (16,6.1527)    (18,5.3990)    (20, 4.7815)    (22,4.2251) (24,3.8082) (26, 3.4117) (28,3.0164)(30,2.4805)(32,2.1747) (34,1.5983) 
			(36,1.3023) (38,1.1135) (40,0.9372)(42,0.7049)(44,0.5035)
		}; 
		
		\end{axis}grid=major,
		\end{tikzpicture}

		\label{llsr}
	}
	\caption{Comparison between the theoretical MAC and the simulated average number of visited nodes in the SD algorithm for BPSK modulation.}
\end{figure}
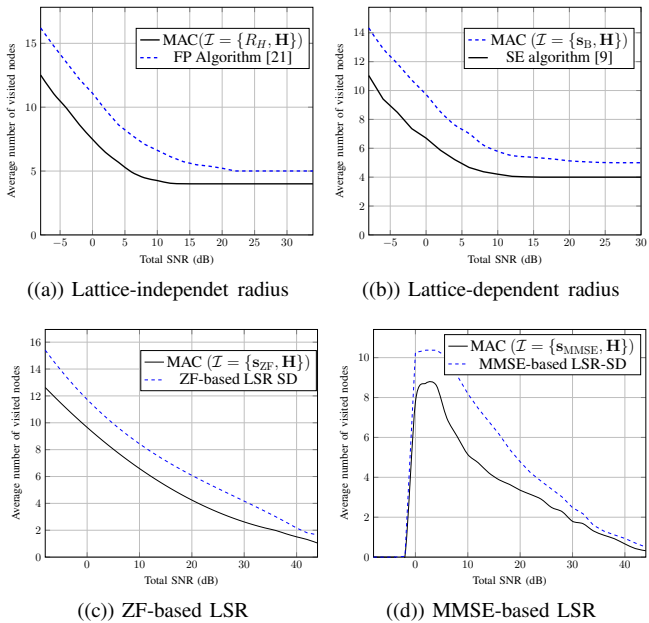
Figs  \ref{fincke}-\ref{llsr} illustrate the MAC  derived in Lemma 1.  The actual FP, SE, and LSR-aided SD methods are also simulated  to compare the number of visited nodes for each method with its corresponding MAC. In this example, we have considered  a MIMO V-BLAST system  with four transmit and receive antennas and BPAM modulation. 
\begin{figure}
	\centering
	\newlength\figureheight 
	\newlength\figurewidth
	\setlength\figureheight{6cm} 
	\setlength\figurewidth{6cm}
	\begin{tikzpicture}[scale=.65]
	\begin{semilogyaxis}[axis background/.style={
		shade,top color=white!10,bottom color=white},
	legend style={fill=white},
	legend style={font=\tiny},
	legend style={at={(0.01,0.01)},
		anchor=south west},
	xmin = -2, xmax = 20,
	ymin = .0000001,
	xlabel={Total SNR (dB)},
	ylabel={Symbol error probability},
	grid=major,
	legend entries={\large MMSE \cite{zeroforcing},\large SD \cite{sphere2017} with $R_\rH$,\large MMSE-based LSR SD, \large WLSD \cite{widely_linear}, \large Threshold (\ref{appeta})},
	]
	
	\addplot+[thick,line join = round,mark =none,color= black]  coordinates{  
		(-2.22,  0.4490   )   ( -.22,   0.3878   )    (1.78,   0.3269   )    (3.78,  0.2639  )   ( 5.78,     0.2050   )    ( 7.78,      0.1515  )      ( 9.78,    
		0.1091    )     (11.78,  0.0721   )    ( 13.78,    0.0484  )    (15.78,   0.0308    )    (17.78,    0.0205    )    (19.78,     0.0129   )   
	};
	\addplot+[thick,mark size =3,smooth,line join = round, color= red, mark=otimes]  coordinates{                                  
		(-2.22,0.4660)   ( -.22,0.4080)    (1.78,0.3398)    (3.78,0.2653)   ( 5.78,0.1577)    ( 7.78,0.0797)      ( 9.78,0.0250)     (11.78,  0.0063)    ( 13.78,8.1667e-04)    (15.78,8.1667e-05)    (17.78,     7.3333e-06)    (19.78,4.6270e-07)     
	};
	\addplot+[thick,mark size =3, line join = round, dashdotted,color= black, mark=diamond]  coordinates{                                  
		(-2.22,0.4660)   ( -.22,0.4080)    (1.78,0.3398)    (3.78,0.2653)   ( 5.78,0.1577)    ( 7.78,0.0797)      ( 9.78,0.0250)     (11.78,  0.0063)    ( 13.78,8.1667e-04)    (15.78,8.1667e-05)    (17.78,     7.3333e-06)    (19.78,4.6270e-07)   
	};
	\addplot+[thick,mark size =3, line join = round,mark=square,color= blue, mark=oplus]  coordinates{                                  
		(-2.22,0.4660)   ( -.22,0.4080)    (1.78,0.3398)    (3.78,0.2653)   ( 5.78,0.1577)    ( 7.78,0.0797)      ( 9.78,0.0297)     (11.78,  0.0075)    ( 13.78,9.4667e-04)    (15.78,10.9667e-05)    (17.78,     11.3333e-06)    (19.78,12.6270e-07)   
	};
	
	\addplot+[thick,mark size =3, line join = round, dashdotted,color= green, mark=oplus]  coordinates{                                  
		(-2.22,0.4660)   ( -.22,0.4080)    (1.78,0.3398)    (3.78,0.2653)   ( 5.78,0.1577)    ( 7.78,0.0797)      ( 9.78,0.0250)     (11.78,  0.0063)    ( 13.78,8.1667e-04)    (15.78,8.1667e-05)    (17.78,     7.3333e-06)    (19.78,4.6270e-07)   
	};
	\end{semilogyaxis}grid=major,

	\end{tikzpicture}

	\caption{ Flat fading MIMO: The symbol error probability comparison for  six transmit and  receive antennas and 4QAM.}
	\label{performanceMIMO}
\end{figure}
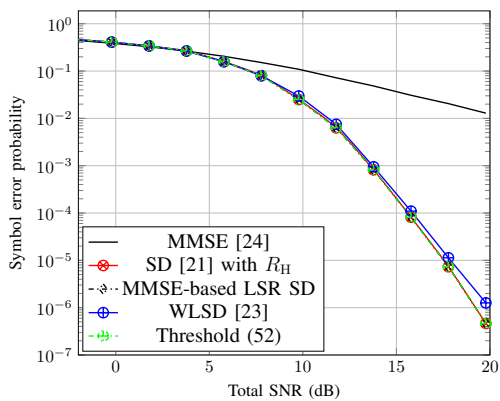
 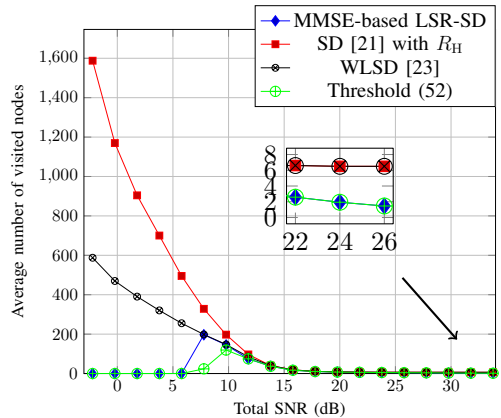
\begin{figure}
 	\centering
 	\setlength\figureheight{3cm} 
 	\setlength\figurewidth{3cm}
 	\begin{tikzpicture}[scale=.65,pin distance=10mm]
 	\tikzstyle{every pin edge}=[<-,thick] 
 	\begin{axis}[axis background/.style={
 		shade,top color=white!10,bottom color=white},
 	legend style={fill=white},
 	legend style={font=\tiny},
 	legend style={at={(1,1.07)},
 		anchor=north east},
 	xlabel={Total SNR (dB)},
 	ylabel={Average number of visited nodes},
 	grid=major,
 	xmin = -3, xmax = 34,
 	ymin = 0,
 	legend entries={\large MMSE-based LSR-SD  , \large SD \cite{sphere2017} with $R_\rH$, \large WLSD \cite{widely_linear}, \large Threshold (\ref{appeta})    },
 	]
 	\addplot+[ thin,color = blue,mark = diamond*,mark size=3]  coordinates{                                                         
 		(-2.22, 0 )   ( -.22,0 )    (1.78,0 )    (3.78,0)   ( 5.78,   0.5580  )    (7.78,   196.7780  )      ( 9.78,   145.1955  )     (11.78,  74.5930   )    ( 13.78,   36.8755 )    (15.78,   16.7770  )    (17.78,    10.0540  )    (19.78,    7.3860  )    (21.78,  6.0400    )    (23.78,  5.1720 )   ( 25.78,     4.3110    )   ( 27.78,   3.4965  )   ( 29.78, 2.5835 )   ( 31.78,     1.9385)  (33.78, 1.4640 )
 	};
 	\addplot+[ thin, mark =square*,color = red]  coordinates{                                                         
 		(-2.22, 1587.6 )   ( -.22,  1169.7 )    (1.78,  904.4 )    (3.78, 700.3 )   ( 5.78, 495.3 )    ( 7.78,  328.3 )    ( 9.78, 197.4 )    ( 11.78,  96.9 )     (13.78,  40.5   )    ( 15.78,  18.5 )    (17.78,  11.2 )    (19.78,    8.4 )    (21.78,   7.3 )    (23.78, 6.9   )      ( 25.78,   6.7   )   ( 27.78,  6.6 )   ( 29.78, 6.6 )   ( 31.78,    6.5 )  ( 33.78,    6.5 )  
 	};
 	\addplot+[ thin, mark =otimes,color = black]  coordinates{                                                         
 		(-2.22, 587.6 )   ( -.22,  469.7 )    (1.78,  390.4 )    (3.78, 320.3 )   ( 5.78, 255.3 )    ( 7.78,  198.3 )    ( 9.78, 147.4 )    ( 11.78,  84.9 )     (13.78,  40.5   )    ( 15.78,  18.5 )    (17.78,  11.2 )    (19.78,    8.4 )    (21.78,   7.3 )    (23.78, 6.9   )      ( 25.78,   6.7   )   ( 27.78,  6.6 )   ( 29.78, 6.6 )   ( 31.78,    6.5 )  ( 33.78,    6.5 )  
 	}; 
 	\addplot+[ thin,color =green,mark = oplus, mark size =3]  coordinates{                                                         
 		(-2.22, 0 )   ( -.22,0 )    (1.78,0 )    (3.78,0.0045)   ( 5.78,   0.0190  )    (7.78,   23.7780  )      ( 9.78,   119.1955  )     (11.78,  74.5930   )    ( 13.78,   36.8755 )    (15.78,   16.7770  )    (17.78,    10.0540  )    (19.78,    7.3860  )    (21.78,  6.0400    )    (23.78,  5.1720 )   ( 25.78,     4.3110    )   ( 27.78,   3.4965  )   ( 29.78, 2.5835 )   ( 31.78,     1.9385)  (33.78, 1.4640 )
 	};
 	\end{axis}grid=major,
 	\node[pin=above left:{%
 		\begin{tikzpicture}[trim axis left,trim axis right]
 		\begin{axis}[
 		xmin=22,xmax=26,
 		ymin=0,ymax=8,
 		line join=round,
 		enlargelimits,width = 3cm
 		]
 		\addplot+[ thin,color = blue,mark = diamond*, mark size=3]  coordinates{                                                         
 			(-10, 0 )   ( -8,0 )    (-6,0 )    (-4,0)   ( -2,   0.5580  )    ( 0,   196.7780  )      ( 2,   145.1955  )     (4,  74.5930   )    ( 6,   36.8755 )    (8,   16.7770  )    (10,    10.0540  )    (12,    7.3860  )    (14,  6.0400    )    (16,  5.1720 )   ( 18,     4.3110    )   ( 20,   3.4965  )   ( 22, 2.5835 )   ( 24,     1.9385)  (26, 1.4640 )
 		};
 		\addplot+[ thin, mark =square*,color = red]  coordinates{                                                         
 			(-10, 1587.6 )   ( -8,  1169.7 )    (-6,  904.4 )    (-4, 700.3 )   ( -2, 495.3 )    ( 0,  328.3 )    ( 2, 197.4 )    ( 4,  96.9 )     (6,  40.5   )    ( 8,  18.5 )    (10,  11.2 )    (12,    8.4 )    (14,   7.3 )    (16, 6.9   )      ( 18,   6.7   )   ( 20,  6.6 )   ( 22, 6.6 )   ( 24,    6.5 )  ( 26,    6.5 )  
 		}; 
 		\addplot+[ thin, mark =otimes,color = black, mark size=3]  coordinates{                                                         
 			(-10, 1587.6 )   ( -8,  1169.7 )    (-6,  904.4 )    (-4, 700.3 )   ( -2, 495.3 )    ( 0,  328.3 )    ( 2, 197.4 )    ( 4,  96.9 )     (6,  40.5   )    ( 8,  18.5 )    (10,  11.2 )    (12,    8.4 )    (14,   7.3 )    (16, 6.9   )      ( 18,   6.7   )   ( 20,  6.6 )   ( 22, 6.6 )   ( 24,    6.5 )  ( 26,    6.5 )  
 		}; 
 		\addplot+[ thin,color = green,mark = oplus,mark size =3]  coordinates{                                                         
 			(-10, 0 )   ( -8,0 )    (-6,0 )    (-4,0)   ( -2,   0.5580  )    ( 0,   196.7780  )      ( 2,   145.1955  )     (4,  74.5930   )    ( 6,   36.8755 )    (8,   16.7770  )    (10,    10.0540  )    (12,    7.3860  )    (14,  6.0400    )    (16,  5.1720 )   ( 18,     4.3110    )   ( 20,   3.4965  )   ( 22, 2.5835 )   ( 24,     1.9385)  (26, 1.4640 )
 		};
 		\end{axis}
 		\end{tikzpicture}%
 	}] at (7.8,0.5) {};
 \end{tikzpicture}

 	\caption{Flat fading MIMO: The complexity  comparison for six transmit and  receive antennas and 4QAM.}
 	\label{complexityMIMO}
 \end{figure}
\begin{figure}
	\centering
	\setlength\figureheight{6cm} 
	\setlength\figurewidth{6cm}
	\begin{tikzpicture}[scale=.65]
	\begin{semilogyaxis}[axis background/.style={
		shade,top color=white!10,bottom color=white},
	legend style={fill=white},
	legend style={font=\tiny},
	legend style={at={(0.01,0.01)},
		anchor=south west},
	xmin = -2, xmax = 26,
	ymin = .00001,
	xlabel={Total SNR (dB)},
	ylabel={Symbol error probability},
	grid=major,
	legend entries={\large MMSE \cite{zeroforcing},\large SD \cite{sphere2017} with $R_\rH$,\large MMSE-based LSR SD, \large WLSD \cite{widely_linear}, \large Threshold (\ref{appeta})},
	]
	
	\addplot+[thick,line join = round,mark =none,color= black]  coordinates{  
		(-4,.8887) (1,.8213) (6,.6938) (11,.5195) (16,.3001)(21,.1300)(26,.0473)  
	};
	\addplot+[thick,mark size =3,smooth,line join = round, color= red, mark=otimes]  coordinates{                                  
		(-4,.8492) (1,.7823) (6,.6605) (11,.4457) (16,.1250)(21,.0073)(26,.0001)       
	};
	\addplot+[thick,mark size =3, line join = round,color=green, mark=diamond,mark size=3]  coordinates{                                  
		(-4,.8492) (1,.7823) (6,.6605) (11,.4457) (16,.1250)(21,.0073)(26,.0001)  
	};
	\addplot+[thick,mark size =3, line join = round,color= blue, mark=oplus, mark size=3]  coordinates{                                  
		(-4,.8492) (1,.7823) (6,.6685) (11,.4487) (16,.1350)(21,.0095)(26,.0002)  
	};
	\addplot+[thick,mark size =3, line join = round,mark=square,color= blue,dashdotted, mark=diamond, mark size=3]  coordinates{                                  
		(-4,.8492) (1,.7823) (6,.6605) (11,.4457) (16,.1250)(21,.0073)(26,.0001)   
	};

	\end{semilogyaxis}grid=major,

	\end{tikzpicture}

	\caption{ Flat fading MIMO: The symbol error probability comparison for  four transmit and  receive antennas and 64QAM.}
	\label{performanceMIMO2}
\end{figure}
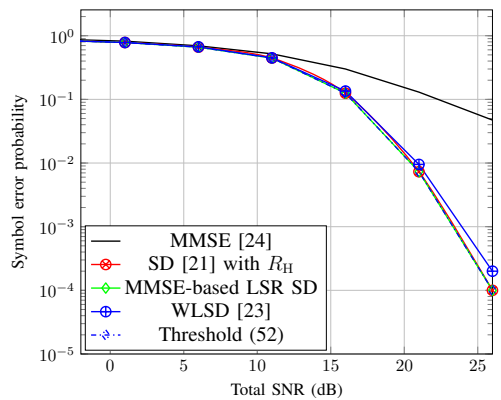 
\begin{figure}
	\centering
	\setlength\figureheight{3cm} 
	\setlength\figurewidth{3cm}
	\begin{tikzpicture}[scale=.65,pin distance=12mm]
	\tikzstyle{every pin edge}=[<-,thick] 
	\begin{axis}[axis background/.style={
		shade,top color=white!10,bottom color=white},
	legend style={fill=white},
	legend style={font=\tiny},
	legend style={at={(.95,1.1)},
		anchor=north east},
	xlabel={Total SNR (dB)},
	ylabel={Average number of visited nodes},
	grid=major,
	xmin = 6, xmax = 26,
	ymin = 0,
	legend entries={\large MMSE-based LSR-SD  , \large SD \cite{sphere2017} with $R_\rH$, \large WLSD \cite{widely_linear}, \large Threshold (\ref{appeta})    },
	]
	\addplot+[ thin,color =blue,mark = oplus, mark size =3]  coordinates{                                                         
		(-4,   0    ) (1,    0    ) (6,   0    ) (11,    353    ) (16,    24    )(21,   9    )(26, 5    )    
	};
	\addplot+[ thin,color = red,mark = square*,mark size=3]  coordinates{                                                         
		(-4, 5643  ) (1,  2665  ) (6, 1462  ) (11, 757  ) (16,  355  )(21, 142  )(26, 84  )  
	};
	\addplot+[ thin, mark =square*,color = black]  coordinates{                                                         
		(-4, 2143  ) (1,  1665  ) (6, 662  ) (11, 357  ) (16,  235  )(21, 142  )(26, 84  )   
	};
	\addplot+[ thin, mark =otimes,color = green]  coordinates{                                                         
		(-4,   0    ) (1,    0    ) (6,   0    ) (11,    53    ) (16,    24    )(21,   9    )(26, 5    )     
	}; 
	\end{axis}grid=major,
	\node[pin=above left:{%
		\begin{tikzpicture}[trim axis left,trim axis right]
		\begin{axis}[
		xmin=21,xmax=26,
		ymin=8,ymax=200,
		line join=round,
		enlargelimits,width = 3.2cm
		]
		\addplot+[ thin,color =blue,mark = oplus, mark size =3]  coordinates{                                                         
			(-10,   0    ) (-5,    0    ) (0,   0    ) (5,    353    ) (10,    24    )(15,   9    )(20, 5    )    
		};
		\addplot+[ thin,color = red,mark = square*,mark size=3]  coordinates{                                                         
			(-10, 5643  ) (-5,  2665  ) (0, 1462  ) (5, 757  ) (10,  355  )(21, 142  )(26, 84  )  
		};
		\addplot+[ thin, mark =square*,color = black]  coordinates{                                                         
			(-10, 2143  ) (-5,  1665  ) (0, 662  ) (5, 357  ) (10,  235  )(21, 142  )(26, 84  )   
		};
		\addplot+[ thin, mark =otimes,color = green]  coordinates{                                                         
			(-10,   0    ) (-5,    0    ) (0,   0    ) (5,    53    ) (10,    24    )(21,   9    )(26, 5    )     
		};
		\end{axis}
		\end{tikzpicture}%
	}] at (7.8,0.5) {};
\end{tikzpicture}

	\caption{Flat fading MIMO: The complexity  comparison for four transmit and  receive antennas and 64QAM.}
	\label{complexityMIMO2}
\end{figure}
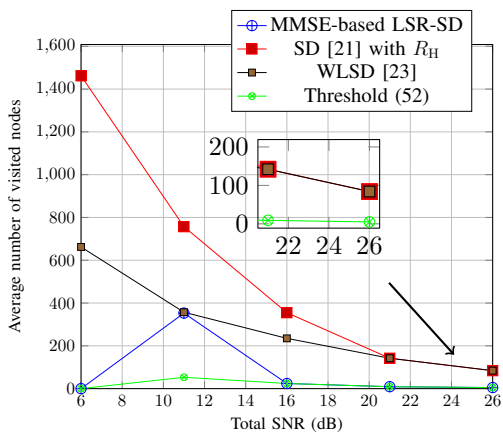 
In order to calculate the MAC  in Figs.~\ref{fincke} and \ref{schnorr}, the information sets, $\mathcal{I}_{\rm FP}=\left\{R_{\rm H},\bH \right\}$ and $\mathcal{I}_{\rm SE}=\left\{\bs_{\rm B},\bH \right\}$ are considered for the SD algorithm where $R_\rH$ is the proposed radius in \cite{hasibi} and $\bs_\rB$ is the Babai point. It can be seen in Figs. \ref{fincke} and \ref{schnorr}, that at  low SNR regime, the MAC of both FP and SE variants of the SD algorithm demonstrates a high computational complexity. 
\begin{figure*}
	\centering
	\subfloat[Total SNR =$-5$dB]{
		\begin{tikzpicture}[scale=.5]
		\begin{axis}[axis background/.style={
			shade,top color=white!10,bottom color=white},
		legend style={fill=white},
		legend style={font=\tiny},
		legend style={at={(0.7,.9)},
			anchor=north east},
		xlabel={Number of antennas},
		ylabel={Average number of visited nodes},
		grid=major,
		xmin = 4, xmax = 16,
		ymin = 0,
		legend entries={\large SD with $R_\rH$  , \large MMSE-based LSR-SD},
		]
		\addplot+[thick,color = red,mark = diamond]  coordinates{                                                         
			(4, 20.968 )     (6,  63.758 )    ( 8,  167.134 )    (10, 450.206 )    (12, 1185.702 )    (14,  3026.87 )    (16,  7687.148 )
		};
		\addplot+[ thick, mark =square,color = blue]  coordinates{                                                         
			(4,   0.064  )     (6,    0.48  )    ( 8,    2.11  )    (10,  3.8639999999999  )    (12,  5.4759999999999  )    (14,   117.534  )    (16,    360.454  ) 
		}; 
		
		\end{axis}grid=major,

		\end{tikzpicture}

		\label{antenna1}	
	}
	\subfloat[Total SNR =$15$dB]{
		\begin{tikzpicture}[scale=.5]
		\begin{axis}[axis background/.style={
			shade,top color=white!10,bottom color=white},
		legend style={fill=white},
		legend style={font=\tiny},
		legend style={at={(0.7,.9)},
			anchor=north east},
		xlabel={Number of antennas},
		ylabel={Average number of visited nodes},
		grid=major,
		xmin = 4, xmax = 16,
		ymin = 0,
		legend entries={\large SD with $R_\rH$  , \large MMSE-based LSR SD},
		]
		\addplot+[thick,color = red,mark = diamond]  coordinates{                                                         
			(4,   6.398999999999954   )     (6,  9.029999999999991    )    ( 8,  11.989999999999958   )    (10,   15.621999999999934    )    (12, 20.332999999999959   )    (14,   26.203999999999862     )    (16,   34.409999999999975     )  
		};
		\addplot+[thick, mark =square,color = blue]  coordinates{                                                         
			(4,   1.976999999999979    )     (6,  4.767999999999960     )    ( 8,   7.303999999999952   )    (10,    9.491999999999957   )    (12,  14.419999999999948    )    (14,     18.660999999999937     )    (16,    25.175999999999934 )   
		}; 
		
		\end{axis}grid=major,

		\end{tikzpicture}

		\label{antenna2}	
	}
	\subfloat[Total SNR =$25$dB]{
		\begin{tikzpicture}[scale=.5]
		\begin{axis}[axis background/.style={
			shade,top color=white!10,bottom color=white},
		legend style={fill=white},
		legend style={font=\tiny},
		legend style={at={(0.7,.9)},
			anchor=north east},
		xlabel={Number of antennas},
		ylabel={Average number of visited nodes},
		grid=major,
		xmin = 4, xmax = 16,
		ymin = 0,
		legend entries={\large SD with $R_\rH$   ,\large MMSE-based LSR-SD},
		]
		\addplot+[thick,color = red,mark = diamond]  coordinates{                                                         
			(4,   4.511000000000001    )     (6,  6.524000000000004 )    ( 8,    8.529000000000009   )    (10,     10.543999999999832    )    (12,   12.580000000000007     )    (14,      14.612999999999795    )    (16,     16.642999999999972  )  
			
		};
		\addplot+[thick, mark =square,color = blue]  coordinates{                                                         
			(4,    0.039000000000000    )     (6,   0.386000000000000  )    ( 8,      0.733000000000000   )    (10,     1.080000000000000   )    (12,  1.427000000000000    )    (14,       1.774000000000000   )    (16,     2.01000000000000  )  
		}; 
		
		\end{axis}grid=major,

		\end{tikzpicture}

		\label{antenna3}
	}
	\caption{Flat fading MIMO: Complexity comparison when the number of antennas increases}
\end{figure*}
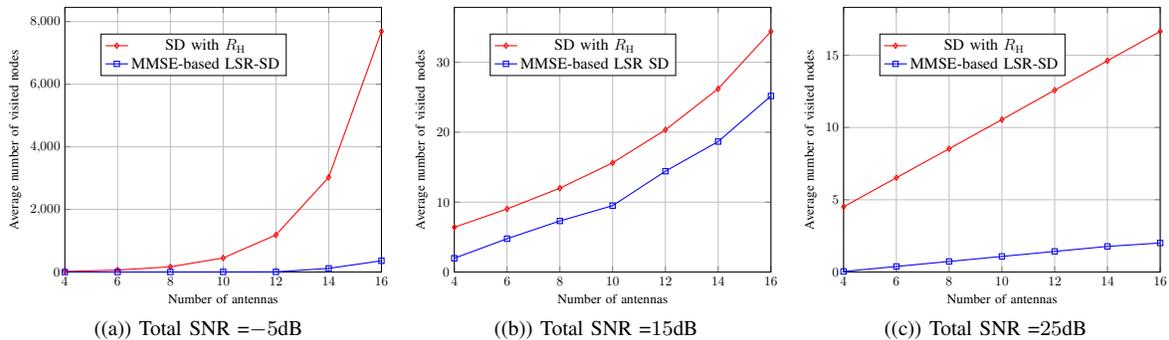
It should be noted that the  gap between the MAC and the actual number of visited nodes decreases with increase of the SNR. Figs \ref{llsr} and \ref{ZFllsr} depict the behavior of the MAC for an MMSE-based LSR and ZF-based LSR with $\mathcal{I}=\left\{\bs_{\rm MMSE},\bH \right\}$ and  $\mathcal{I}=\left\{\bs_{\rm ZF},\bH \right\}$, respectively.  This behavior confirms  the ZCA property of the MMSE-based LSR at low and high SNR regimes. Nonetheless, it can be seen that adding the ZF detected symbol to the information set does not lead to ZCA property at low SNRs. In other words, as it can be seen in Lemma 2, the ZF detected point  is not able to resolve the low SNR deficiency of the SD algorithm.   This confirms the fact that the ZF-based LSR-aided SD does not obey  marginal complexity optimality. 
\subsection{Frequency flat Rayleigh Fading MIMO Channel}
  In this simulation example, the MIMO channels are considered to be i.i.d complex Gaussian with zero mean and unit variance.  The performance of the proposed method is evaluated through symbol error probability. The number of visited nodes  in the SD algorithm is also calculated in order to compare the complexity of the proposed method with that of some other detection methods.   
 
 In Fig. \ref{performanceMIMO}, a MIMO system with six transmit and receive antennas and 4-ary QAM is considered, and it can be seen that  the proposed method achieves the ML performance. The same configuration  is considered in Fig. \ref{complexityMIMO}, to  compare the complexity of the proposed method with that of the  SD  algorithm presented in \cite{sphere2017} with $R_\rH$. For comparison, the performance and the computational complexity of the widely linear SD in \cite{widely_linear} is also plotted. It can be seen that  the proposed algorithm achieves a lower complexity and, unlike the other SD algorithms, is as complex as the MMSE algorithm at low and high SNRs. This  result corroborates our claims in Theorems 2 and 3, which means that the proposed algorithm achieves the ML performance at all SNRs and exploits the ZCA property. Moreover, it can be seen that the approximate threshold proposed in (\ref{appeta}), with a scaling factor of two, also leads to a performance which is very close to that of the ML detector. It also demonstrates a significantly lower computational complexity. The same results can be observed from Figs. \ref{performanceMIMO2} and \ref{complexityMIMO2} for 64-ary QAM and four receive and transmit antennas. It can be seen that the performance threshold proposed in (\ref{appeta}) is very close to the ML detector while achieving a significantly lower computational complexity.     

 Figs~\ref{antenna1}-\ref{antenna2} and \ref{antenna3} demonstrate the effect of increasing the number of antennas on the complexity of the MMSE based LSR-SD algorithm for $\rho_T=-5,15$, and $25$. It can be observed that the LSR  in the MMSE based LSR-SD compensates the SD algorithm for its high computational complexity, especially at low SNRs. Although at high SNRs, the SD algorithm has almost a linear complexity increase with the number of antennas, the MMSE-based LSR-SD leads to a dramatic complexity decrease.
 \vspace*{-4mm} 
\subsection{Frequency Selective Channel}
\vspace*{-1mm} 
To evaluate the performance of the  MMSE based LSR-SD algorithm in another application, in this simulation example, we consider  a wireless point to point frequency selective channel with Rayleigh fading. The $L_c$ channel coefficients $\bh = [h_0 \dots,h_{L_c-1}]$ are i.i.d  distributed with zero mean and unit variance. The channel input-output model is 
\beq \label{fs}
y[n] = \sum_{l=0}^{L_c-1}h_l s[n-l]+w[n],
\eeq
where the additive noise $w[n]$ is i.i.d with zero mean and variance $\sigma^2$. The transmission scheme is considered to be single carrier zero padding (ZP) block transmission \cite{tseisi}. One can reformulate (\ref{fs}) as the system model (\ref{systemmodel}) \cite{tseisi}. In this example, the number of channel coeficients is considered to be $L_c=7$, the  transmit vector lengths is $K=8$, and consequently, the receive vector length is $L=K+L_c-1$.  
\begin{figure}
	\centering
	\setlength\figureheight{6cm} 
	\setlength\figurewidth{6cm}
	\begin{tikzpicture}[scale=.65]
	\begin{semilogyaxis}[axis background/.style={
		shade,top color=white!10,bottom color=white},
	legend style={fill=white},
	legend style={font=\tiny},
	legend style={at={(0.03,0.03)},
		anchor=south west},
	xmin = 9, xmax = 17,
	ymin = .00001,
	xlabel={Total SNR (dB)},
	ylabel={Symbol error probability},
	grid=major,
	legend entries={\large ZF \cite{tseisi}, \large SD \cite{sphere2017} with $R_\rH$, \large MMSE-based LSR SD, \large Threshold (\ref{appeta})},
	]
	
	\addplot+[thick,line join = round,mark=square,color= black]  coordinates{  
		(9,0.4441)    (11,0.2363)    (13,0.0745)   (15,0.0107)    (17,0.0007)     
	};;
	\addplot+[thick,mark size =3,smooth,line join = round, color= red, mark=otimes]  coordinates{                                  
		(9,0.3879)    (11,0.1768)    (13,0.0339)    (15,0.0017) (17,.000067678)    
	};
	\addplot+[thick,mark size =3,smooth,line join = round, color= blue, mark=o]  coordinates{                                  
		(9,0.3879)    (11,0.1768)    (13,0.0339)    (15,0.0017)    (17,.000067678)
	};
	\addplot+[thick,mark size =3,smooth,line join = round, color= green,dashdotted, mark=o]  coordinates{                                  
		(9,0.3879)    (11,0.1968)    (13,0.0339)    (15,0.0017)    (17,.000067678)
	};
	\end{semilogyaxis}grid=major,
	\end{tikzpicture}

	\caption{Frequency selective: The symbol error probability comparison $L_c=7$, $K=8$ and $L=14$ for 16QAM.}
	\label{performancefs}
\end{figure}
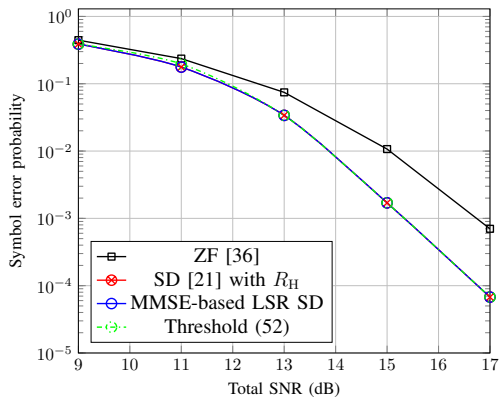
\begin{figure}
	\centering
	\setlength\figureheight{6cm} 
	\setlength\figurewidth{6cm}
	\begin{tikzpicture}[scale=.65,pin distance=10mm]
	\tikzstyle{every pin edge}=[<-,thick]
	\begin{axis}[axis background/.style={
		shade,top color=white!10,bottom color=white},
	legend style={fill=white},
	legend style={font=\tiny},
	legend style={at={(0.4,0.8)},
		anchor=south west},
	xlabel={Total SNR (dB)},
	ylabel={Number of visited nodes},
	grid=major,
	xmin = 7, xmax = 21,
	ymin = 0,
	legend entries={\large MMSE-based LSR-SD  ,\large  SD \cite{sphere2017} with $R_\rH$, \large Threshold (\ref{appeta})},
	]
	\addplot+[thick,color = blue,mark = diamond*]  coordinates{                                                         
		(7,  0 )   ( 9,  116.4847 )    (11,  31.0936 )    ( 13, 9.9342    )     (15,   4.0074   )     (17, 1.4071  )    (19, 0.3482     )   (21,0.0754 ) 
	};

	\addplot+[thick,mark size =3,line join = round,mark=otimes,color= red]  coordinates{ 
		(7,   1048.9  )   ( 9,   174.7  )    (11,   40.7  )    ( 13,   11   )     (15,    8.1  )     (17, 8  )    (19,8    )   (21,8 )   
	}; 
	\addplot+[thick,color = green,mark = oplus, mark size=3]  coordinates{                                                         
		(7,  0 )   ( 9,  50.4847 )    (11,  27.0936 )    ( 13, 9.9342    )     (15,   4.0074   )     (17, 1.4071  )    (19, 0.3482     )   (21,0.0754 ) 
	};
	\end{axis}grid=major,
	\node[pin=above left:{%
		\begin{tikzpicture}[trim axis left,trim axis right]
		\begin{axis}[
		xmin=8,xmax=12,
		ymin=0,ymax=10,
		line join=round,
		enlargelimits,width = 3.1cm
		]
		\addplot+[thick,color = blue,mark = diamond*]  coordinates{                                                         
			(-2,  0 )   ( 0,  116.4847 )    (2,  31.0936 )    ( 4, 9.9342    )     (6,   4.0074   )     (8, 1.4071  )    (10, 0.3482     )   (12,0.0754 ) 
		};
		\addplot+[thick,mark size =3,line join = round,mark=otimes,color= red ]  coordinates{ 
			(-2,   1048.9  )   ( 0,   174.7  )    (2,   40.7  )    ( 4,   11   )     (6,    8.1  )     (8, 8  )    (10,8    )   (12,8 )   
		}; 
		\addplot+[thick,color =green,mark = oplus,mark size=3]  coordinates{                                                         
			(-2,  0 )   ( 0,  116.4847 )    (2,  31.0936 )    ( 4, 9.9342    )     (6,   4.0074   )     (8, 1.4071  )    (10, 0.3482     )   (12,0.0754 ) 
		};
		\end{axis}
		\end{tikzpicture}%
	}] at (7.8,0.5) {};
	
\end{tikzpicture}

	\caption{Frequency selective: The complexity comparison $L_c=7$, $K=8$ and $L=14$ for 16QAM.}
	\label{complexityfs}
\end{figure}
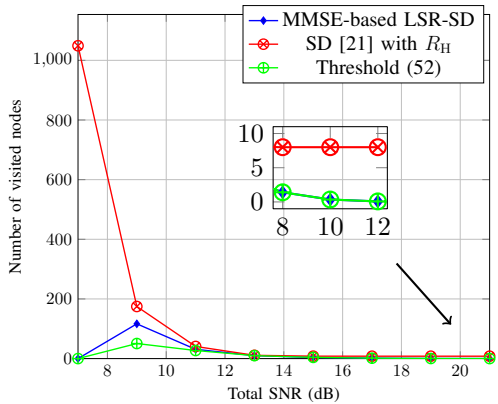
 Fig. \ref{performancefs} demonstrates the symbol error probability for a single carrier ZP system  which adopts 16-QAM as the modulation scheme. Similar to other simulation examples, it is observed that the proposed algorithm  achieves ML performance and outperforms the ZF detector. Also, the approximated threshold (\ref{appeta}), with a scaling factor of two, achieves a very close to ML performance.     

In Fig. \ref{complexityfs}, we compare the complexity of the proposed scheme with other simulated schemes in terms of the number of visited nodes in the reduced SD method. It can be observed that the  MMSE-based LSR-SD algorithm reduces the number of visited nodes in comparison with the SD algorithm and exploits the ZCA property. Moreover, the approximated threshold (\ref{appeta}) leads to a significantly low computational complexity.  
\vspace{-2mm}
\section{Conclusion} \label{conclusion}
Using the law of large numbers, the minimum achievable complexity has been derived for any arbitrary SD algorithm with a set of auxiliary information. The theoretical reason for  a fundamental flaw of the conventional SD algorithms at low SNRs has been analyzed. The lossless size reduction concept has been introduced as a potential solution to the low SNR deficiency of conventional SD algorithms. An ML performance-achieving method which is marginally optimal in the sense of computational complexity has been driven. Along with achieving the exact ML performance and being marginally optimal, the proposed method has been shown to reduce the computational complexity by performing a lossless search over a reduced search space.    
\appendices
\section{The proof of lemma 1} \label{lemma1proof}
To obtain the MAC, we consider a sequence of $N$ ML search candidates as
\beq
\bS_{\rm ML}^k = [\bs^k_{\rm ML}[1],\bs^k_{\rm ML}[2],\dots,\bs^k_{\rm ML}[N]]^T,
\eeq
where $\bs^k_{\rm ML}[n]$ is the $k$ dimensional ML search candidate at the $n$th time slot. In order  to obtain the possible sequences $\bS_{\rm ML}^k$ and  the typical set, the probability $\rP(\bs_{\rm ML}^k|\mathcal{I})$ should be calculated. It is given as
 \beq \label{I}
 \rP \left(\bS_{\rm ML}^k\left| \mathcal{I}\right. \right) = \prod_{n=1}^N \rP \left(\bs_{\rm ML}^k[n]\left|\mathcal{I}\right. \right). 
 \eeq 
 The above equality comes from the statistical independence of $\bs_{\rm ML}^k[n]$'s in different time slots.  Taking the logarithm of both sides of (\ref{I}) yields
\beq 
\frac{1}{N} \log_M \rP \left(\bS_{\rm ML}^k\left| \mathcal{I}\right. \right) = \frac{1}{N}\sum_{n=1}^N \log_M \rP \left(\bs_{\rm ML}^k[n]\left| \mathcal{I}\right. \right). 
 \eeq 
 Applying the law of large numbers, leads to
 \begin{align}
 &\lim\limits_{N\rightarrow \infty} \frac{1}{N}\sum_{n=1}^N \log_M \rP \left(\bs_{\rm  ML}^k[n]\left| \mathcal{I}\right. \right) \nonumber \\
 &= \rE_{\bs_{\rm ML}^k[n]}\left\{ \log_M \rP \left(\bs_{\rm ML}^k[n]\left| \mathcal{I}\right. \right)  \right\}.
 \end{align}
 According to Shannon's AEP, for a given $\mathcal{I}$, the number of typical $k$ dimensional lattice points is 
$
 M^{-\rE_{\bs_{\rm ML}}\left\{    \log_M\rP\left(\bs_{\rm ML}^k|\mathcal{I}\right)\right\} }.
$
Therefore, the number of typical $k$ dimensional lattices is obtained as
$
\rE_{\mathcal{I}} \left\{ M^{-\rE_{\bs_{\rm ML}}\left\{    \log_M\rP\left(\bs_{\rm ML}^k|\mathcal{I}\right)\right\} } \right\}.
$
It should be noted that $|\mathcal{T}_k|$ is the number of typical $k$ dimensional lattices at the $k$th stage of the SD algorithm. Hence, the MAC of an SD algorithm  given the information set $\mathcal{I}$ is     
\beq
\rC^{min}_{\rm SD} =  \rE_{\mathcal{I}} \left\{\sum_{k=1}^K\mathcal{T}_k\right\} = \rE_{\mathcal{I}} \left\{\sum_{k=1}^KM^{-\rE_{\bs^k_{\rm ML}}\left\{ \log\rP\left(\bs_{\rm ML}^k|\mathcal{I}\right)\right\}}\right\}.
\eeq
\vspace*{-4mm}
\section{Proof of lemma 2}\label{lemma2proof}
  In \cite{tsebook}, for the low SNR regime, it is shown that unlike the ZF, the MMSE detected symbol is information lossless.To make further clarification for the low SNRs, it should be noted that we have $\bH\bs = \sum_{k=1}^K\bh_k[\bs]_k$, and the $i$th detected symbol by the ML detector is obtained as 
  \beq
  [{\bs}_{\rm ML}]_i = [\arg \min_{[\bs]_i}\|\by - [\bs]_i\bh_i-\sum_{k=1,k\neq i}^K[\bs]_k\bh_k  \|^2]_i.
  \eeq  
At a low SNR regime, the interference term $\sum_{k=1,k\neq i}^K [\bs]_k\bh_k $ is negligible, and  we can write
\beq \label{64}
  [{\bs}_{\rm ML}]_i = [\arg \min_{[\bs]_i}\|\by - [\bs]_i\bh_i\|^2]_i.
  \eeq
On the other hand, for the maximum ratio combiner (MRC) detector, we have \cite{tsebook} 
\beq \label{65}
[{\bs}_{\rm MRC}]_i = [\arg \min_{[\bs]_i}\|\by - \sqrt{\rho}[\bs]_i\bh_i\|]_i = \rq\left(\frac{[\bH^H\by]_i}{\|\bh_i\|^2} \right),
\eeq
  where $\rq(\cdot)$ is the operation of quantizing to the nearest modulation point. The MMSE detected symbol is
  \beq
  [{\bs}_{\rm MMSE}]_i = {\rq}\left(\frac{1}{\gamma_i}\left[ \left(\bH^H\bH +\frac{1}{\rho}\bI\right)^{-1}\bH^H\by\right]_i\right),
  \eeq
  where
  $
  \gamma_i = 1- \left[ \left(\rho\bH^H\bH +\bI\right)^{-1}\right]_{ii}.
    $ 
   Using the Taylor equation, we have 
  \beq \label{taylor}
  \left(\bH^H\bH +\frac{1}{\rho}\bI\right)^{-1}\approx \rho\left(\bI -\rho\bH^H\bH \right)\by \approx \rho \bH^H \by,
  \eeq
  and $\gamma_i \approx \rho \|\bh_i\|^2$.
 Therefore, for small SNRs,  we have
 \beq \label{68}
 \lim \limits_{\rho\rightarrow 0} [{\bs}_{\rm MMSE}]_i=\rq\left(\frac{[\bH^H\by]_i}{\|\bh_i\|^2} \right) = [{\bs}_{\rm MRC}]_i.
 \eeq
  According to (\ref{64}), (\ref{65}), and (\ref{68}), we have $\lim \limits_{\rho\rightarrow 0}{\bs}_{\rm ML}={\bs}_{\rm MMSE}$. Therefore, 
  \beq
 \lim \limits_{\rho\rightarrow 0} \rP(K_r=0) = \lim \limits_{\rho\rightarrow 0}  \rP\left({\bs}_{\rm ML}={\bs}_{\rm MMSE}\right) =1,
  \eeq
  and  according to (\ref{BB}) for an MMSE-based LSR-aided SD, we have $\lim \limits_{\rho\rightarrow 0}\rC^{min}_{\rm LSR} =0$.
 The ZCA property does not hold for the ZF detector.
   Indeed, for the ZF method, we have  
  $
  [{\bs}_{\rm ZF}]_i = \rq \left([\left(\bH^H\bH\right)^{-1}\bH^H\by]_i\right),
  $
  which in general is not equal to $[{\bs}_{\rm MRC}]_i$, unless the channel matrix $\bH$ is orthogonal. Since for a fading channel, this event occurs with zero probability, a ZF based LSR-aided SD is not guaranteed to be ZCA at $\rho=0$. 

Now we consider the high SNR regime. Using the fact that $\lim \limits_{\rho \rightarrow \infty} \rP({\bs}_{\rm ML} = \bs_{\rm MMSE})=1$, we have $\rP(K_r=0)=1$, and according to (\ref{BB}), we obtain $\lim \limits_{\rho \rightarrow \infty}\rC^{min}_{\rm LSR} = 0$. 

\end{document}